\documentclass[a4paper,11pt]{article}
\usepackage{jcappub}
\usepackage{amsmath}
\usepackage{amssymb}
\usepackage[T1]{fontenc}
\usepackage{siunitx}
\usepackage{subcaption}
\usepackage{listings}
\usepackage[normalem]{ulem}

\newcommand{\nb}{n_\text{b}}

\newcommand{\Rb}{R_\text{b}}

\title{\boldmath General relativistic bubble growth in cosmological phase transitions }

\author[a]{L. Giombi,}
\author[a,b]{Mark Hindmarsh}

\affiliation[a]{Department of Physics and Helsinki Institute of Physics,\\PL 64, FI-00014 University of Helsinki, Finland}
\affiliation[b]{Department of Physics and Astronomy, University of Sussex,\\Brighton BN1 9QH, United Kingdom}

\emailAdd{lorenzo.giombi@helsinki.fi}
\emailAdd{mark.hindmarsh@helsinki.fi}

\abstract{ We use a full general relativistic framework to study the self-similar expansion of bubbles of the stable phase into a flat Friedmann-Lemaître-Robertson-Walker Universe in a first order phase transition in the early Universe.  With a simple linear barotropic equation of state in both phases, and in the limit of a phase boundary of negligible width, we find that self-similar solutions exist, which are qualitatively similar to the analogous solutions in Minkowski space, but with distinguishing features. Rarefaction waves extend to the centre of the bubble, while spatial sections near the centre of the bubble have negative curvature. 
Gravitational effects redistribute the kinetic energy of the fluid around the bubble, and can change the kinetic energy fraction significantly. The kinetic energy fraction of the gravitating solution can be enhanced over the analogous Minkowski solution by as much as $\mathcal{O}(1)$, and suppressed by a factor as larger as $\mathcal{O}(10)$ in case of fast detonations.
The amount of negative spatial curvature at the centre of the bubble is of the same order of magnitude of the naive expectation based on considerations of the energy density perturbation in Minkowski solutions, with gravitating deflagrations less negatively curved, and detonations more. 
We infer that general relativistic effects might have a significant impact on  
accurate calculations of the gravitational wave power spectrum when the bubble size becomes comparable to the cosmological Hubble radius,
affecting the primary generation from the fluid shear stress, and inducing secondary generation by scalar perturbations. 
}

\begin{document}
\maketitle
\flushbottom

\section{Introduction}\label{sec:introduction}
Phase transitions are characteristic phenomena predicted by many gauge field theories. They naturally arise as a consequence of a spontaneous breaking of a gauge symmetry~\cite{Kirzhnits:1972iw, Kirzhnits:1972ut, Kirzhnits:1976ts}. The standard Hot Big Bang model predicts at least two transitions~\cite{Kibble:1980mv, Kolb:1990vq}, one at the electroweak scale $(\sim\SI{100}{\giga\eV})$ and one at the scale of quark confinement $(\sim\SI{100}{\mega\eV})$. In the Standard Model of particle physics, these transitions complete with a crossover~\cite{Kajantie:1996mn, Borsanyi:2016ksw}. First order phase transitions are possible in many extensions of the Standard Model~\cite{Caprini:2019egz}. The characteristic feature of a first order phase transition (FOPT) is the presence of a metastable quasi-equilibrium state just below the critical temperature. Quantum and thermal fluctuations push the field across a potential barrier and initiate the production of bubbles of the stable equilibrium phase, which eventually will expand and merge converting larger and larger portions of the Universe to the stable phase~\cite{Witten:1980ez, Guth:1981uk, Steinhardt:1980wx, Coleman:1977py, Linde:1981zj}.

FOPTs enjoy great interest as a necessary component of electroweak baryogenesis~\cite{Cohen:1990it, Cline:2006ts, Cline:2018fuq}, and the possibility of observing the stochastic gravitational wave background~\cite{Witten:1984rs, Hogan:1986qda, Kosowsky:1992rz, Kamionkowski:1993fg} they produce. 
The scenario for phase transitions becomes particularly compelling if a  FOPT took place around the electroweak scale, because the gravitational wave signal that this transition is expected to generate might lie in the frequency window of future gravitational wave detectors as LISA (Laser Interferometer Space Antenna)~\cite{LISA:2017pwj}. 

The large energy density perturbations created by a strong phase transition might also seed a population of primordial black holes~\cite{Kodama:1982sf,Hawking:1982ga,Khlopov:1998nm,Franciolini:2021nvv,Liu:2021svg,Lewicki:2023ioy}, 
which in turn would produce gravitational waves when they become gravitationally bound (see e.g.~\cite{Sasaki:2018dmp} for a review, also \cite{Bartolo:2018rku, Franciolini:2022ewd, LISACosmologyWorkingGroup:2022jok}). Gravitational wave production by primordial black holes has been under intense investigation since the first direct detection of gravitational waves from binary black holes merging by the Virgo-LIGO collaboration~\cite{LIGOScientific:2016aoc, LIGOScientific:2016sjg, LIGOScientific:2017bnn} in 2015.

Predictions of the gravitational wave power spectrum directly produced at a  FOPT~\cite{Caprini:2015zlo, Caprini:2019egz, Hindmarsh:2020hop, Guo:2020grp} indicate that the estimated energy density in the gravitational waves $\Omega_{\textnormal{GW}}$ is an increasing function of $(R_\star H_e)$, with $R_\star$ the mean bubble spacing and $H_e$ the Hubble constant at some convenient reference time during the transition which we take to be the time when a fraction $1/e$ of the Universe remains in the metastable phase. In a conventional thermal phase transition, this is also the time of peak bubble nucleation rate, sometimes written $H_n$~\cite{Enqvist:1991xw}. One is naturally led to consider the possible scenarios that maximize this factor, as these would be the transitions most likely to be observable.
In this paper we focus on the study of the hydrodynamics of large bubbles, by which we mean bubbles of the stable equilibrium phase whose radial size $R_b$ is comparable to the Hubble radius, $R_b\sim H^{-1}$. This possibility is allowed if the nucleation rate of bubble of the stable phase is sufficiently small to allow the bubbles to expand for enough time before merging.
We will see that gravitation affects the fluid velocity and energy density profiles around the bubbles, which will lead to changes in the gravitational wave power spectrum~\cite{Hindmarsh:2020hop}.

Furthermore, work on gravitational wave production from FOPTs has until now considered only  
primary gravitational waves, sourced by the transverse and traceless part of the energy momentum tensor~\cite{Hindmarsh:2020hop, Guo:2020grp, Caprini:2015zlo, Caprini:2019egz}. Secondary gravitational waves sourced by scalar perturbations have been neglected, because they are expected to provide a negligible contribution as long as the bubbles are smaller than the cosmological horizon. However, in slow transitions, if the bubbles' size  become comparable with the Hubble radius, we estimate that scalar induced (secondary) gravitational waves can play an important role. In light of these considerations, we are motivated to study the limit of large bubbles  ($R_b H\sim \mathcal{O}(1)$), that require a full general relativistic treatment.

Right after the nucleation, a pressure difference is established in the plasma filling the Universe at the interface between the symmetric and broken phases.  Once the bubble is set in motion by the pressure difference at the interface, the friction  between the bubble and the particles of the plasma induces a negative acceleration that slows down the expansion of the bubble~\cite{Linde:1981zj, Linde:1980tt}. The competition between the friction and the driving force induced by the pressure difference determines whether, after a sufficient period of time, the bubble reaches a steady state with a constant terminal velocity, or continue to accelerate~\cite{Moore:1995si, Liu:1992tn, John:2000zq, Bodeker:2017cim}. Steady-state solutions are also possible without entropy generation~\cite{Enqvist:1991xw, Moore:1995si, Ai:2021kak}. Phase transition fronts that expand at subsonic speed have been shown to be unstable to perturbations with wavelength $\lambda_c\lesssim\lambda \lesssim R_b$, where $\lambda_c$ is a critical length scale set by the surface tension and the kinetic energy difference across them~\cite{Link:1992dm}. The effect of non-spherical perturbations is to increase the bubble phase boundary and thereby accelerate the bubble expansion~\cite{Kamionkowski:1992dc} to supersonic speeds. We leave  investigation of instabilities for future work.

On the macroscopic scale the cosmic plasma around the bubble admits a fluid description. Numerical analysis of the fluid profile of the bubble~\cite{Espinosa:2010hh, Hindmarsh:2020hop} as well as numerical simulations of FOPTs~\cite{Hindmarsh:2013xza, Hindmarsh:2015qta, Hindmarsh:2017gnf} have been carried on a flat background, and the effect of the expansion of the Universe has only been included as a rescaling of the energy-momentum tensor of the radiation fluid. 
This approximation relies of the fact that, if the transition completes in a period of time that is much shorter than the Hubble time, then the bubble size will remain much smaller then the Hubble radius, and, in a first approximation, the resulting gravitational waves behave as if in Minkowski space, after a conformal rescaling.
As we said, this approximation breaks in slow phase transitions when the bubble size become comparable to the Hubble radius. The hydrodynamics of bubble in an expanding Universe described by a flat Friedmann-Lemaître-Robertson-Walker (FLRW) solution were studied in~\cite{Cai:2018teh}. However, the back-reaction of the bubble expansion on the bulk FLRW metric was neglected. 

In this work we focus on the expansion of a single bubble during its steady state evolution and we wish to go beyond the approximation of Ref.~\cite{Cai:2018teh} considering back-reaction between the spacetime metric and the bubble. Full general relativity has already been applied to study the collision between two vacuum bubbles ~\cite{Johnson:2011wt}. Here we wish to gain 
a full general relativistic understanding of the fluid dynamics around a single bubble, which will be important for accurate predictions of both primary and secondary gravitational wave production in slow phase transitions.

We find convenient to use geometrized units in this paper, where the Newton constant $G$ and the speed of light $c$ are set to unity, $G = c =1$.

\section{Phase transitions in the Early Universe}\label{sec:Phase_transition_early_Universe} 
The content of the early Universe before the phase transition can be described on large scales as a perfect fluid in thermal equilibrium with energy momentum
\begin{equation}\label{def:T_f}
    T_{\mu\nu}^f = wu_\mu u_\nu + p g_{\mu\nu},
\end{equation}
with the enthalpy density $w = e+p$, where $e$ and $p$ are the energy density and pressure respectively; $u^\mu$ is the four-velocity of the fluid, and $g_{\mu\nu}$ the background metric. The simplest way to model a system with a first order phase transition is to add a scalar field $\phi$ coupled to the plasma which acts as an order parameter for the transition. The energy momentum tensor of the full scalar-fluid system is then~\cite{Kurki-Suonio:1995yaf}
\begin{equation}\label{def:T_phi}
    T_{\mu\nu} = wu_\mu u_\nu + p g_{\mu\nu} +\nabla_\mu \phi \nabla_\nu \phi -  \frac{1}{2}g_{\mu\nu} \left(\nabla\phi\right)^2 ,
\end{equation}
where $w,\, e$ and $p$ depend on both the temperature $T$ and the scalar field $\phi$. We suppose that $\phi = 0$ in the high-temperature phase.

\subsection{Free energy and effective potential}\label{sec:Free_energy_effective_potential}
All the thermodynamic properties of the scalar-fluid system are encoded in the pressure $p(\phi,T)$, or equivalently in the free energy density $\mathcal{F}(\phi,T) = - p(\phi,T)$~\cite{Hindmarsh:2016lnk, Hindmarsh:2019phv, Hindmarsh:2020hop}. Given the free energy, one can identify the potential of the zero temperature theory $V_0(\phi)$ and the effective potential $V_{\textnormal{eff}}(\phi, T)$, arising from thermal corrections and interactions between the scalar field $\phi$ and the particles in the plasma as
\begin{eqnarray}
    V_0(\phi) &\equiv& \mathcal{F}(\phi, 0),\\
    V_{\textnormal{eff}}(\phi,T) &\equiv& \mathcal{F}(\phi,T) - \mathcal{F}(0,T).
\end{eqnarray} 
Both the energy and enthalpy density derive from the free energy according to~\cite{Cai:2018teh}
\begin{equation}
    w = T\frac{\partial \mathcal{F}}{\partial T}, \qquad\qquad e = \mathcal{F} - T\frac{\partial \mathcal{F}}{\partial T}.
\end{equation}
At any temperature $T$, the equilibrium states of the system are found as the states of minimal free energy\footnote{Notice that the equilibrium states of the free energy are also minima of the effective potential.}
\begin{equation}
    \frac{\partial\mathcal{F}(\phi,T)}{\partial \phi} = 0, \qquad \frac{\partial^2\mathcal{F}(\phi,T)}{\partial \phi^2} > 0.
\end{equation}
In the general scenario~\cite{Hindmarsh:2020hop}, at high temperature there is only one global minimum of the free energy, which uniquely defines the equilibrium state of the system. As the Universe expands, the temperature drops, and the free energy features a new local minimum. At the critical temperature $T_c$, the two minima become degenerate in the free energy density $\mathcal{F}$. Below the critical temperature the system is  initially in a metastable quasi-equilibrium state (sometimes called \textit{false vacuum}). An effective potential barrier separates states with homogeneous $\phi$ and prevents the system from classically rolling to the new stable equilibrium state (sometimes called \textit{true vacuum}), which is now the global minimum for the free energy density. Nonetheless, quantum and thermal fluctuations allow the scalar field $\phi$ to jump through the barrier via inhomogeneous configurations and to reach the new equilibrium state.

Since the transition between the two equilibrium state is sometimes accompanied by a (spontaneous) symmetry breaking, it is common to refer to the system in the false vacuum as \textit{symmetric phase}, and the system in the true vacuum as \textit{broken phase}. We will adopt here the same nomenclature, even though we do not require any symmetry breaking to take place. After nucleation, the bubbles whose radius exceeds a critical value $R_c$~\cite{Kirzhnits:1976ts}, start growing until they eventually merge and and the stable phase percolates.

The bubble wall separates two distinct phases with different equations of state. In the interior of the bubble the scalar-fluid system is in the broken phase, while outside the bubble the system is in the old symmetric metastable phase. Between the two phases there is a \textit{wall}, that is a region where the order parameter $\phi$ smoothly interpolates between the values in the two vacuum states, over a microscopic distance of order $T_c^{-1}$. The wall itself carries a surface tension $\sigma$ that is determined by the effective potential, and is of order $T_c^{3}$.

We mentioned in the Introduction ~\ref{sec:introduction} that our focus is on large bubbles, that is bubble whose broken phase has a radial size $\Rb$ comparable with the Hubble radius $H^{-1}$. As the bubble grows large, we are less and less interested in resolving the microphysics at the wall. Denoting with $W$ the proper width of the wall, the difference in scale $W/\Rb \ll 1$ motivates us to adopt the \textit{thin wall approximation}, that is to consider the wall as an infinitesimally thin surface, and to treat the order parameter $\phi$ as a step function with a discontinuity at the wall.

\subsection{Equation of state}\label{sec:eos}
After nucleation, the pressure difference at the interface between the two phases forces the bubble to expand. To describe this process we need an equation of state (EOS) that links the thermodynamic variables $p(\phi, T), \, e(\phi,T)$. For the purpose of this paper, we will consider a simple scenario where the scalar-fluid system behaves as a barotropic perfect fluid in both phases, with equation of state
\begin{equation}
p = \omega e
\end{equation}
where $\omega$ is the EOS parameter, taken to be constant. For simplicity we will fix the EOS parameter in the broken phase to $\omega_- = 1/3$, and  consider different values of the EOS parameter $\omega_+$ in the symmetric phase. We further assume that $\omega_\pm$ evolves slowly with the temperature $T$ throughout the phase transition. For numerical calculations, this allows us to consider, in a first approximation, the EOS parameter $\omega_\pm$ to be constant during the expansion of the bubble. In order for the bubble to expand we need $\omega_+ < \omega_-$; this condition reflects the fact that some particle species acquire a mass in the broken phase and become non-relativistic. Qualitatively speaking, the number of particle species that become non-relativistic in the broken phase determines the strength of the phase transition. To quantify this concept, it is useful to introduce the trace anomaly $\theta$
\begin{equation}\label{def:theta}
    \theta \equiv \frac{1}{4}(e-3p). 
\end{equation}
We can then define the boundary strength parameter of the phase transition $\alpha_+$ as~\cite{Hindmarsh:2020hop} 
\begin{equation}\label{def:alpha_plus}
    \alpha_+ = \frac{4}{3}\frac{\theta_+ - \theta_-}{w_+}.
\end{equation}
where the plus (minus) sign indicates that the quantity must be evaluated just ahead (behind) the interface. 
We also introduce the related idea of the global strength parameter 
\begin{equation}\label{def:alpha_n}
    \alpha = \frac{4}{3}\frac{\theta_s - \theta_b}{w_s}\,
\end{equation}
where all quantities are evaluated at the same temperature, typically that at which the bubble nucleates, and the subscripts $s$ or $b$ indicates that the quantity is evaluated in the symmetric or broken phase. 
The two quantities are not equal in general, as the temperature in front of the wall can be higher than the nucleation temperature, and is different from the temperature behind the wall. However, in our simplified equation of state, the quantities $\theta$ and $w$ scale the same way with temperature, and $\alpha_+ = \alpha$.

\subsection{Mean bubble separation}\label{sec:mean_bubble_separation}
Once the temperature of the Universe drops below the critical temperature $T_c$, spherical bubbles of the new phase start nucleating from thermal and quantum fluctuations of the field $\phi$. Due to the random nature of quantum and thermal fluctuations, the appearance of bubbles of the new broken phase is a stochastic process that can happen everywhere in the Universe. After nucleation, the bubbles start growing until 
they eventually merge and percolate, and the  Universe converts to the new phase. We introduce a new length scale $R_\star$ which defines the mean bubble centre spacing, so that
\begin{equation}
R_* = [\nb(t\rightarrow \infty)]^{-1/3},
\end{equation}
where $\nb$ is the number density of bubbles.
The length scale $R_\star$ is important for observations, since it sets the peak frequency of the gravitational waves generated by the phase transition~\cite{Hindmarsh:2016lnk, Hindmarsh:2019phv, Hindmarsh:2020hop, Caprini:2015zlo, Caprini:2019egz, Weir:2017wfa}. The interest of this work is in the limit of large bubbles ($R_\star H\sim \mathcal{O}(1)$), which requires a full general relativistic treatment.

\subsection{Bubble growth in Minkowski spacetime}\label{sec:mink}
\label{sec:bubble_expansion}

In this section we briefly review the results on bubble dynamics on a Minkowski background. This might help the reader to establish a comparison with the general relativistic treatment of the following sections.

The bubble wall is the interface between the symmetric and broken phases, 
and below the critical temperature there is a pressure difference across the wall, 
which drives the expansion of the bubble. The microscopic description of the expansion is given by a solution of the effective field equation for the scalar field $\phi$~\cite{Kurki-Suonio:1995yaf}
\begin{gather}\label{eq:KG}
    \Box \phi - \frac{\partial V_{\textnormal{eff}}}{\partial \phi} = \eta(\phi,T) u^\mu\partial_\mu\phi,
\end{gather}
where $\eta(\phi,T)$ encapsulates the microscopic out-of-equilibrium dynamics leading to friction between the expanding bubble and the particles of the plasma~\cite{Moore:1995si, Liu:1992tn, Krajewski:2023clt}. The friction function $\eta(\phi,T)$ is difficult to calculate and is often treated as a free constant parameter~\cite{Kurki-Suonio:1995yaf}. 
Its main function is to determine the asymptotic speed of the wall as the bubble grows large. The steady flow solution is usually reached after a first period of accelerated expansion; once the wall reaches its terminal velocity, the bubble continues to expand at constant speed until colliding with other bubbles in the Universe. In this paper we will focus on describing bubbles during their steady flow state. 

Equation~\eqref{eq:KG} is non trivial only at the wall, where the field $\phi$ interpolates between its values in the broken and symmetric phase. By the equivalence principle, there must be a system of coordinates that is comoving with the wall and in which the metric is locally Minkowski; we call this system the \textit{Gaussian normal frame}~\cite{Israel:1966rt}. Therefore, in Gaussian normal coordinates the coupled scalar field - fluid equation~\eqref{eq:KG} is the same in the general relativistic case and in the flat Minkowski approximation, and so should be relationship between the friction $\eta(\phi,T)$ and the fluid speed in the wall frame should also be the same~\cite{Kurki-Suonio:1995yaf}.

Let us consider a flat static Universe. In polar coordinates $x^\mu = (t,r,\theta,\phi)$, the metric is
\begin{equation}
    ds^2 = \eta_{\mu\nu}dx^\mu dx^\nu, \qquad \eta_{\mu\nu} = \textnormal{diag}(-1,1,r^2,r^2\sin^2\theta),
\end{equation}
and the fluid four-velocity
\begin{equation}
    u^\mu = \frac{dx^\mu}{\sqrt{-ds^2}} = \gamma\left(1,u,0,0\right),\qquad \gamma_u \equiv \frac{1}{\sqrt{1-u^2}}
\end{equation}
with $u \equiv dr/dt$ the outward fluid three-velocity. Once the bubble reaches its steady state, the wall moves with a constant velocity $u_w$ with respect to the Universe rest frame (\textit{Universe frame}). In the rest frame of the wall (\textit{wall frame}) the conservation of the energy-momentum at the wall implies
\begin{subequations}\label{eq:junction_Mink}
\begin{eqnarray}
    \left[T^{rt}\right]^\pm &=& 0 \quad \rightarrow\quad \left[w\gamma^2 v\right]^\pm = 0,\\
    \left[T^{rr}\right]^\pm &=& 0 \quad \rightarrow\quad \left[w\gamma^2 v^2 + p\right]^\pm = 0,
\end{eqnarray}
\end{subequations}
where $v$ is inward fluid velocity in the wall frame\footnote{$v>0$ means that the fluid in entering the bubble. The bubble then sweeps over the symmetric phase.} and $\gamma = (1-v^2)^{-1/2}$; for convenience we introduced the bracket notation 
\begin{subequations}\label{def:brackets}
\begin{eqnarray}
    \left[A\right]^\pm \equiv A_+ - A_-,\\
    \left\{A\right\}^\pm \equiv A_+ + A_- .
\end{eqnarray}
\end{subequations}
The subscript positive (negative) sign convention, consistently with equation~\eqref{def:alpha_plus}, denotes quantities evaluated just outside (inside) the interface.  Equations~\eqref{eq:junction_Mink} admit an analytical solution $v_\pm(\alpha_+, v_\mp)$
\begin{subequations}\label{eq:v_pm}
\begin{eqnarray}
    v_+ &=& \frac{1}{2(1+\alpha_+)} \left[\frac{1}{3v_-} + v_- \pm \sqrt{\left(\frac{1}{3v_-} - v_-\right)^2 + \frac{8}{3}\alpha_+ + 4\alpha_+^2 }\right],\label{eq:v_+}\\
    v_- &=& \frac{1}{2} \left[\left(\left(1+\alpha_+\right)v_+ + \frac{1-3\alpha_+}{3v_+}\right) \pm \sqrt{\left(\left(1+\alpha_+\right)v_+ + \frac{1-3\alpha_+}{3v_+}\right)^2 -\frac{4}{3}}\right],\label{eq:v_-}
\end{eqnarray}
\end{subequations}
for a given strength parameter $\alpha_+$~\eqref{def:alpha_plus}. Physical solutions of~\eqref{eq:v_+} demands the positive sign when $v_- > 1/3$ and the negative sign when $v_- \leq 1/3$. Analogously, physical solutions of~\eqref{eq:v_-} demand the positive sign for $v_+ < 1/3$ and negative sign for $v_+ \geq 1/3$. This means that both velocities inside and outside the wall can be either subsonic (deflagration) or supersonic (detonation). The strength parameter $\alpha_+$ for subsonic solutions has an upper bound, $\alpha_+ < 1/3$, given by the requirement that $v_+$ must be positive. The choice of sign in~\eqref{eq:v_pm} depends on the wall speed in the Universe rest frame and the sound speed $c_s$ of the broken phase
\begin{equation}
    c_s^2 \equiv \frac{dp\left(\phi(T),T\right)}{dT} \left(\frac{de\left(\phi(T),T\right)}{dT}\right)^{-1},
\end{equation}
where $\phi(T)$ is the equilibrium value of the field.
As stated in Section~\ref{sec:eos}, we suppose that the effective potential is such that the scalar-fluid system is a barotropic perfect fluid in thermal equilibrium, so that the pressure is a function of the energy density only, $p=p(e)$, and we compute the speed of sound as 
\begin{equation}
    c_s^2 = \frac{d p(e)}{de}.
\end{equation}
We will comment more about the sign choices in~\eqref{eq:v_pm} later. 

The profiles of the fluid velocity and enthalpy density as functions of $r$ and $t$ are found by projecting the conservation equations onto the fluid four-velocity $u^\mu$ and its spacelike orthogonal vector $\Bar{u}^\mu \equiv \gamma_u(-u, 1, 0, 0)$:
\begin{equation}
    0 = u_\mu\nabla_\nu T^{\mu\nu} = \bar{u}_\mu\nabla_\nu T^{\mu\nu}.
\end{equation}
Let us now choose our time coordinate $t$ such that the bubble nucleates at $t=0$. Since the bubble is expanding on a flat background, there is no physical length scale involved in the problem. We then expect the solution of conservation equations to manifest a self-similar behaviour. This means that the physical variables of the system, namely the fluid three-velocity and enthalpy density, depend on the radial and time coordinates only through the combination $\xi\equiv r/t$. For the purpose of numerical calculation, it is convenient to write the system of equations of motion in a parametric form, adding a new parameter $\tau$ such that
\begin{subequations}\label{eq_motion_mink}
\begin{eqnarray}
    \frac{d\xi}{d\tau} &=& \xi\left[(\xi- u)^2 -c_s^2\left(1-\xi u\right)^{2}\right],\\
    \frac{du}{d\tau} &=& 2vc_s^2(1-u^2)(1-\xi u),\\
     \frac{dw}{d\tau} &=& w \left(1+\frac{1}{c_s^2}\right) \gamma^2 \mu \frac{du}{d\tau} 
\end{eqnarray}
\end{subequations}
where we have defined
\begin{equation}\label{def:mu}
    \mu(u,\xi) \equiv  \frac{\xi - u}{1-\xi u}
\end{equation}
the outward fluid velocity in a frame that is moving outward at speed $\xi$. The system~\eqref{eq_motion_mink} has three fixed points in the phase space $(\xi, u)$; one at the origin $(0,0)$, one on the light cone at $(1,1)$ and one at $(c_s,0)$. Technically speaking, the point $(c_s, 0)$ is an unstable improper node~\cite{Espinosa:2010hh}, and all the trajectories approach it tangentially to the axis $(\xi, u=0)$.

For a given wall speed $u_w$, the system~\eqref{eq_motion_mink} uniquely provide the bubble profile once we impose boundary conditions at the origin of coordinates, at $\xi = u_w$, and at infinity. At $\xi=0$ we expect $u = 0$ because of the spherically symmetric configuration. On the other hand, $\xi=1$ represent the position of the light cone. Since information cannot travel faster than light, we also expect the velocity field to reach zero before the light cone, that is $u(\xi=1)=0$. In between we use the junction conditions 
\eqref{eq:junction_Mink} to join interior and exterior solutions.

\begin{figure}
    \centering
    \includegraphics[width=\textwidth]{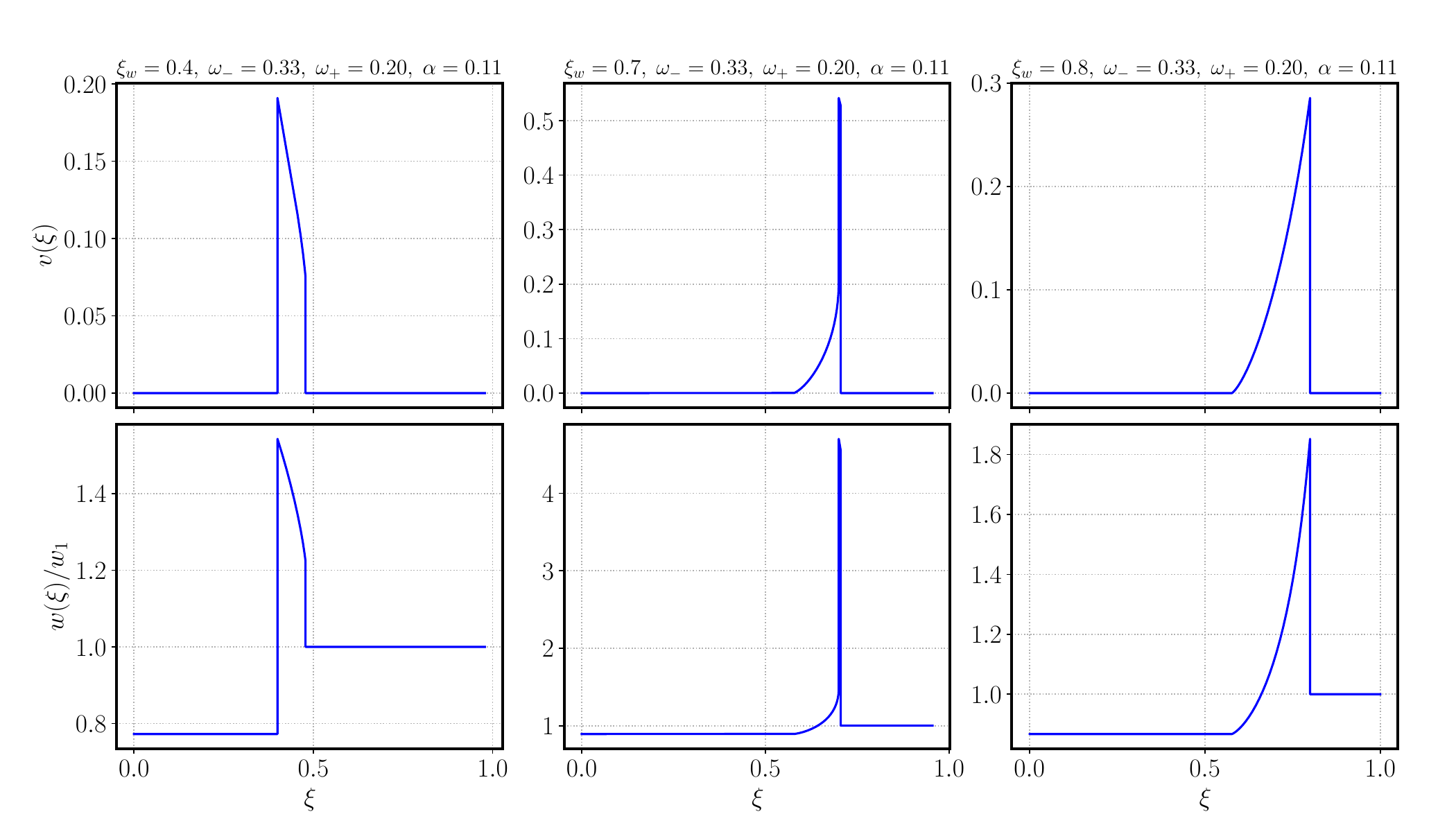}
    \caption{Profiles of outward fluid velocity $v$ (top panels) and enthalpy $w$ normalized to its value outside the shock $w_1$ (bottom panels) against the self-similar coordinate $\xi=r/t$ with the constant sound speed equation of state in the case of a bubble expanding on a flat Minkowski background for deflagration (left panels), hybrid (central panels) and detonation (right panels). In the constant sound speed model $\omega_-$ ($\omega_+$) is the speed of sound squared in the broken (symmetric) phase, and $\alpha$ is the transition strength parameter \eqref{def:alpha_n}, which is temperature-independent in this model.}  
    \label{fig:mink}
\end{figure}

Depending on the way we meet the boundary conditions, we distinguish three cases:
\begin{itemize}
    \item \textit{Subsonic deflagration:} The bubble expands in a deflagration mode when the inward fluid velocity in the wall frame behind the wall is slower than the sound speed of the broken phase, $v_- < {c_s}_-$. The fluid is at rest inside the wall in the Universe frame, hence $u_w = v_-$.
    We must choose the negative sign in~\eqref{eq:v_+} in order to ensure that the fluid flows into the bubble, i.e. $u_+ = \mu(v_+,\xi_w) >0$. In front of the wall, the fluid velocity profile shows a compression wave that decays until the fluid reaches, via a shock, the configuration at rest with the Universe frame ($u=0$).

     \item \textit{Detonation:} A detonation is characterized by a supersonic inward fluid exit speed in the wall frame $v_-> {c_s}_-$, while the fluid is at rest with the Universe frame everywhere outside the wall $u_+ = \mu(v_+,\xi_w) =0$. The velocity profile shows a rarefaction wave behind the wall that reaches the improper node at $({c_s}_-, 0)$. 
     The condition $v_->{c_s}_-$ is satisfied choosing the positive sign in  equation~\eqref{eq:v_-}, and it implies that there is a minimum possible value for $v_+$ given by $v_\textnormal{CJ} \equiv v_+(\alpha_+, {c_s}_-)$. This is the Chapman-Jouguet speed; in our case\footnote{This is also the case, for example, of the Bag equation of state~\cite{Hindmarsh:2019phv, Cai:2018teh}.} we have ${c_s}_- =1/\sqrt{3}$, for which the Chapman-Jouguet speed is 
     \begin{equation}
         v_\textnormal{CJ} = \frac{1}{\sqrt{3}}\left( \frac{1+\sqrt{\alpha_+ + 3\alpha_+^2}}{1+\alpha_+}\right).
     \end{equation}
     Therefore in a detonation $u_w>v_{\textnormal{CJ}}$.
    
    \item \textit{Supersonic deflagration (hybrid):} A hybrid is a solution at wall speeds in between the two aforementioned cases; more precisely, in a hybrid solution the inward fluid exit speed in the wall frame is precisely $v_- = {c_s}_-$, and the wall speed in the Universe frame is bounded by ${c_s}_- <u_w < v_\textnormal{CJ}$. The velocity profile shows both a rarefaction and a compression wave behind and outside the wall respectively. The rarefaction wave ends at the fixed point $({c_s}_-, 0)$ as in a detonation, while the compression wave ends with a shock as in a deflagration. At the wall, we choose the negative sign in~\eqref{eq:v_+} in order to ensure $u_+ >0$. Outside the shock the fluid is at rest with the Universe frame $u^{sh}_+=0$. 
    Hybrids may not be stable solutions for some equations of state~\cite{Krajewski:2023clt}.
\end{itemize}
The solutions for the three different cases have been studied and discussed in~\cite{Kurki-Suonio:1995yaf,Espinosa:2010hh,Hindmarsh:2019phv}. In Figure~\ref{fig:mink} we show an example of fluid velocity and enthalpy profiles where the sound speed $c_s^2 \equiv dp/de$ is constant in the two phases (see Eq.~\eqref{def:eos} and~\eqref{def:omega}) for a mathematical definition of the EOS). Plots were produced using a modification of the code snippet provided in Ref.~\cite{Giese:2020znk}. In the following we will use this class of Minkowski spacetime solution with constant sound speed EOS as a reference to compare to the gravitating bubble solutions.

\section{General relativistic description of bubble growth} \label{sec:General_relativistic_bubble_growth}
\subsection{Coordinate frame and equations of motion}\label{sec:coordinate_frame}
We will consider that, during the expansion of the bubble, both the inner and outer regions are described by a spherically symmetric spacetime. We parametrize the expanding region as a foliation of concentric spherical shells labeled with a radial coordinate $r$ comoving with the fluid~\cite{Musco:2012au}. The metric is
\begin{equation}\label{eq:metric}
    ds^2 = -a^2 dt^2 + b^2 dr^2 +R^2 \left( d\theta^2 +\sin^2\theta d\varphi^2\right),
\end{equation}
with $a, b$ and $R$ functions of $t$ and $r$ only. This metric was previously studied in the context of gravitational collapse~\cite{Misner:1964je}, droplets in the quark-hadron transition~\cite{Rezzolla:1995kv}, and primordial black hole formation~\cite{Musco:2012au}.  As the radial coordinate is comoving with the fluid, the radial component of the fluid $4$-velocity is zero, as discussed in~\cite{Deng:2017uwc}. In particular
\begin{equation}\label{def:fluid}
    u^\mu = \frac{1}{a}\delta^{\mu 0}.
\end{equation}
The time coordinate can be fixed by specifying the metric coefficient $a$ on a timelike curve~\cite{Deng:2017uwc}. We chose to set $a(0,t) = 1$, so that the cosmic time $t$ refers to the time measured by an observer at the centre of the bubble. We discussed in Section~\ref{sec:bubble_expansion} that steady state solutions with constant wall speed always exist in Minkowski spacetime. 
Defining a wall speed in general relativity can be more difficult, 
and we rather characterize steady state solutions as bubbles whose wall is located at $R_w = \xi_w t$, with $\xi_w$ a constant. 
The position of the wall in comoving coordinates $r_w(t)$ is determined implicitly from  $R(r_w(t),t) =\xi_w t$. We will see that the solutions we find have constant fluid speeds at the wall, similar to those in the Minkowski solutions.

Using the notation of~\cite{Misner:1964je}, we define the proper time and radial derivatives
\begin{equation}
    D_t \equiv u^\mu\partial_\mu = \frac{1}{a}\partial_t, \qquad D_r = \Bar{u}^\mu \partial_\mu = \frac{1}{b}\partial_r,
\end{equation}
where $\Bar{u}^\mu$ is a radial spacelike vector orthogonal to the fluid four-velocity $u^\mu$, and we introduce the quantities
\begin{equation}
    U \equiv D_t R, \qquad \Gamma \equiv D_r R.
\end{equation}
The former represents the radial component of the fluid four-velocity in a (non-comoving) \textit{Eulerian} frame where $R$ is the radial coordinate~\cite{Musco:2018rwt} (hence $R= \sqrt{A/4\pi}$, with $A$ the constant-$r$ spatial surface, is sometimes called the Schwarzschild radial coordinate). The quantity $\Gamma$ 
can be interpreted as a generalization of the fluid Lorentz factor. 
Let us consider the invariant quantity $\Delta$, defined as
\begin{equation}\label{def:delta}
    \Delta \equiv \partial_\mu R \partial^\mu R = -U^2 +\Gamma^2 \equiv 1-2\frac{M}{R}  
\end{equation}
which in turn defines a coordinate invariant quantity $M$ that we will interpret as mass.

Using the Einstein equations we find
\begin{equation}\label{eq:drM}
    \partial_r M = 4\pi e R^2\partial_r R ,
\end{equation}
which gives to $M$ the interpretation of the mass-energy contained inside a shell of radius $R$. Finally we introduce two new dimensionless quantities, the surface energy $\Omega$ on a shell of radius $R$, and the gravitational potential $\Phi$ at $R$ as 
\begin{equation} \label{def:omega-phi}
    \Omega \equiv 4\pi e R^2, \qquad \Phi \equiv \frac{M}{R}.
\end{equation}
In terms of $U, \Omega, \Phi$ the system of Einstein and hydrodynamic equations is~\cite{Musco:2012au, Musco:2018rwt} 
\begin{subequations}\label{equations_motion}
\begin{eqnarray}
D_t U &=& -\left[\dfrac{\Gamma}{e+p}D_rp +\dfrac{\Phi}{R}+\omega\dfrac{\Omega}{R} \right]\\
 D_t\Phi &=& - \dfrac{U}{R}(\Phi + \omega \Omega) \label{eq:D_tPhi}\\
 D_r\Phi &=& -\dfrac{\Gamma}{R}(\Phi -  \Omega) \label{eq:D_rPhi}\\
 D_tb &=& \dfrac{b}{\Gamma}D_rU\\ 
 D_ra &=& \dfrac{a}{U}D_t\Gamma\\
 D_t\Omega &=& 2\dfrac{U\Omega}{R}  -4\pi\dfrac{e+p}{\Gamma}D_r\left(U R^2 \right)\\
 D_ra &=& -\dfrac{a}{(e+p)}D_rp
\end{eqnarray}
\end{subequations}
To close the system, we require an equation of state $p = p(e)$ . Remembering the discussion in Section~\ref{sec:eos} we consider the simple EOS
\begin{equation}\label{def:eos}
    p = \omega(t,r) e, \qquad \frac{d p}{d e}  = c_s^2.
\end{equation}
with $\omega(t,r)$ a constant, but different, value in the two phases. 
The time dependence in $\omega(t,r)$ is caused by the expansion of bubble wall into the symmetric phase. We can write the EOS parameter $\omega$ as a distribution defined on the whole spacetime as
\begin{equation}\label{def:omega}
    \omega(t,r) = \omega_- \Theta(r_w(t)-r) + \omega_+ \Theta(r-r_w(t))
\end{equation}
with $\Theta(r)$ the Heaviside distribution and $r_w(t)$ the position of the wall in comoving radial coordinate  at time $t$. Analogously, the speed of sound can be expressed as 
\begin{equation}\label{def:c_s_dist}
    c_s^2(t,r) = {c^2_s}_- \Theta(r_w(t)-r) + {c^2_s}_+ \Theta(r-r_w(t)).
\end{equation}
This equation of state is a special case of the constant sound speed model \cite{Giese:2020znk}, with zero vacuum energy in both phases. 
For simplicity we fix $\omega_- = 1/3$ in the broken phase, while the EOS in the symmetric phase $\omega_+$ is a free constant parameter.
Fixing $\omega_- =1/3$ implies that the boundary transition strength parameter $\alpha_+$ is independent of the temperature, and given by 
\begin{equation}\label{eq:alpha_plus_us}
    \alpha_+ = \frac{1-3\omega_+}{3(1+\omega_+)}.
\end{equation}
The temperature-independence, and the fact that $\theta_- = 0$, also means that the global transition strength parameter $\alpha = \alpha_+$. Therefore we do not need to make a distinction between the two in this 
special equation of state.

\subsection{Self-similar assumption}\label{sec:self_similar_assumtpion}
Motivated by the results in Minkowski space (see Section~\ref{sec:mink}), we seek self-similar solutions of the physical dimensionless quantities $U(\xi)$, $\Omega(\xi)$, and $\Phi(\xi)$. Self-similar collapsing solutions are already known in the study of primordial black hole formation~\cite{Musco:2012au}. We here adopt a similar formulation and introduce the self-similar variable\footnote{Note that in Ref.~\cite{Musco:2012au} the self-similar variable $\xi$ is defined with the opposite sign, i.e. $\xi = -R/t$, as appropriate for collapsing solutions. }
\begin{equation}\label{def:xi}
    \xi \equiv \frac{R(t,r)}{t}.
\end{equation}
which greatly simplifies the set of Einstein equations to a system of three differential equations
\begin{subequations}\label{equations_motion_xi}
\begin{eqnarray}
    \dfrac{d\ln  U}{d\ln\xi} &=& \left[ (\Phi+\omega\Omega)^2-2c_s^2\Gamma^2\Phi \right]\left[\dfrac{\Omega - \Phi}{U^2(\Phi+\omega\Omega)^2 - c_s^2\Gamma^2(\Omega-\Phi)^2}\right],\label{eq_motion_U}\\
    \dfrac{d\ln \Omega}{d \ln\xi} &=&  \dfrac{\Omega - \Phi}{\Phi+\omega\Omega}\left[ 2\omega + (1+\omega)\dfrac{d \ln U}{d\ln\xi}\right],\label{eq_motion_Omega}\\
    \dfrac{d \ln \Phi}{d\ln\xi} &=& \dfrac{1}{\Phi} \left( \Omega - \Phi\right),\label{eq_motion_Phi}
\end{eqnarray}
where $\Gamma(\xi)$ is given by
\begin{equation}
    \Gamma^2 = 1 + U^2 - 2\Phi.
\end{equation}
\end{subequations}
We notice that the system~\eqref{equations_motion_xi} is autonomous and enjoys the shift symmetry $\ln \xi\rightarrow \ln\xi + \lambda $, with $\lambda$ an arbitrary constant. 
Furthermore, introducing the self-similar variable and combining equations~\eqref{eq:D_tPhi} and~\eqref{eq:D_rPhi}, we find that the metric coefficient $a(t,r)$ also has a self-similar behavior given by
\begin{equation}\label{def:a_ξ}
    a(\xi) = \frac{\Omega- \Phi}{U\Omega (1+\omega)}\xi.
\end{equation}

The junction conditions in Minkowski are stated in the frame at rest with the wall (see Section~\ref{sec:mink}). Analogously we want to define here the notion of a fluid three-velocity for some observer moving outward with the wall. As the wall is assumed to be stationary at a fixed position $\xi_w$, we find convenient to introduce a set of constant-$\xi$ observers, that is observers that are moving outwards with respect to the fluid rest frame with constant $\xi$, defined by the condition $d\xi/dt = 0$. The worldline of these observers in the $(t,r)$ coordinates is described by a 4-velocity \begin{equation}\label{def:const_xi}
    V_\xi^\mu = \gamma\left(\frac{1}{a}, \frac{v}{b}, 0, 0\right)
\end{equation}
with
\begin{equation}\label{def:v}
    v \equiv -\frac{U-\xi/a}{\Gamma}, \qquad \gamma^2 =\frac{1}{1-v^2}.
\end{equation}
We can re-express the definition~\eqref{def:v} of $v$ using the expression~\eqref{def:a_ξ} for $a$ as
\begin{equation}\label{eq:v_expression}
    v = \frac{U}{\Gamma}\frac{\Phi+\omega\Omega}{\Omega-\Phi}.
\end{equation}
This is a more convenient expression to use to evaluate junction conditions and to find critical points.

The relative speed $v_\textnormal{rel}$ between two observers moving with 4-velocities $u_1^\mu$ and $u_2^\mu$ can be computed as shown in~\cite{2002GReGr..34.1491G}
\begin{equation}
    v_\textnormal{rel}^2(u_1,u_2) = \frac{(u_1 \cdot n_2)^2}{(u_1 \cdot n_2)^2 + 1},
\end{equation}
with $n_2^\mu$ the radial orthonormal vector to $u_2^\mu$, 
and $a \cdot b = g_{\mu\nu} a^\mu a^\nu$. One can verify that
\begin{equation}
    v_\textnormal{rel}^2(u,V_\xi) = v^2
\end{equation}
with $u^\mu$ the fluid 4-velocity~\eqref{def:fluid} and $V_\xi^\mu$ the 4-velocity of a constant-$\xi$ observer~\eqref{def:const_xi}. 
Hence we interpret the 3-velocity $v$ as the inward fluid velocity for a constant-$\xi$ observer. At $\xi=\xi_w$, constant-$\xi$ observers are comoving with the wall, and  $v(\xi_w)$ is the inward fluid speed in the wall rest frame.

The simplification from the system~\eqref{equations_motion} of partial differential equations to the system ~\eqref{equations_motion_xi} of three ordinary differential equations comes with some assumptions. First of all, we implicitly assumed that the bubble nucleates at $t^\prime = 0$. Otherwise, one should define the self-similar variable as $\xi = R/(t-t^\prime)$, and explicit time dependent factors $t^\prime/t$ would appear in the system~\eqref{equations_motion_xi}, spoiling the self-similar behavior. The EOS parameter $\omega$ defined in~\eqref{def:omega} is consistent with $H\propto t^{-1}$. Steady state solutions $R_w = \xi_w t$ imply $R_w H\sim \xi_w$. Therefore, self-similar steady state solutions of~\eqref{equations_motion_xi} represent bubbles whose radial size is a constant fraction, approximately given by $\xi_w$, of the Hubble radius $H^{-1}$. In other words, self-similarity is compatible with a steady flow. However, in a real phase transition the EOS parameter is expected to decreases as the Universe cools down. Therefore, the two pictures are compatible if we assume that $\omega$ evolves slowly in time and the time-dependent solutions evolve slowly with $\omega$.

\subsection{Kinematic and geometric properties of an expanding bubble}
The kinematic properties of fluid worldlines encode important information about the behavior of the congruence of the fluid four-velocity $u^\mu$. These properties can be read from the irreducible tensorial decomposition of the covariant derivative of the fluid four-velocity~\cite{2013rehy.book.....R, Andersson:2006nr}
\begin{equation}
    \nabla_\nu u_\mu = \omega_{\mu\nu} + \sigma_{\mu\nu} + \frac{1}{3}\theta \tilde{h}_{\mu\nu} - a_\mu u_\nu
\end{equation}
with $a^\mu \equiv u^\nu\nabla_\nu u^\mu$ the fluid four-acceleration and $\Tilde{h}_{\mu\nu} = g_{\mu\nu} + u_\mu u_\nu$ the projector onto the hypersurface orthogonal to $u^\mu$, and
\begin{equation}
    \omega_{\mu\nu} \equiv \tilde{h}^\alpha_{\phantom{\alpha}\mu}\tilde{h}^\beta_{\phantom{\beta}\nu}\nabla_{[\beta}u_{\alpha]}, \qquad \sigma_{\mu\nu} = \nabla_{(\mu}u_{\nu)} + a_{(\mu}u_{\nu)} - \frac{1}{3}\theta \tilde{h}_{\mu\nu}, \qquad \theta = \tilde{h}^{\mu\nu}\nabla_\nu u_\mu
\end{equation}
are the vorticity tensor, shear tensor, and local expansion scalar respectively. Vorticity describes a rotation of the congruences around a fixed axis; a non vanishing shear introduces distortion on the congruences without changing the volume of spacetime they cover; the expansion scalar instead describes the volume changing of the congruences. We compute these quantities in the case of a spherically symmetric expanding bubble. Vorticity is always zero around a single bubble, while for the shear and expansion scalar we have respectively
\begin{eqnarray}
    \sigma &\equiv& \sqrt{\frac{1}{2}\sigma^{\mu\nu}\sigma_{\mu\nu}}= \frac{U}{\sqrt{3}R} \left\vert 1-\frac{d\ln U}{d\ln\xi} \right\vert, \\
    \theta &=& \frac{U}{R}\left(2+  \frac{d\ln U}{d\ln\xi}\right).
\end{eqnarray}
In a FLRW Universe, $U = RH$ and $d\ln U/d\ln\xi = 1$, so that $\sigma =0$ and $\theta = 3H$.

The curvature properties of the spacetime can be described by the mean of coordinate invariant contractions of the Riemann and Weyl tensor~\cite{Carroll:2004st}. For the purposes of this paper we will be interested only in the Ricci curvature scalar $R^{(4)}$ and Kretschmann scalar $K^{(4)}$
\begin{subequations}
    \begin{eqnarray}
        R^{(4)} &\equiv& R_{\mu\nu\phantom{\rho}\sigma}^{\phantom{\mu\nu}\rho} g^\mu_{\phantom{\mu}\rho} g^{\nu\sigma},\\
        K^{(4)} &\equiv& R_{\mu\nu\rho\sigma}R^{\mu\nu\rho\sigma},
    \end{eqnarray}
\end{subequations}
with $R_{\mu\nu\phantom{\rho}\sigma}^{\phantom{\mu\nu}\rho}$ the Riemann tensor. Using the Einstein equations~\eqref{equations_motion}, after some algebra we find
\begin{subequations}
    \begin{eqnarray}
         R^{(4)} &=& \frac{2}{R^2}(1-3\omega)\Omega ,\\
        K^{(4)} &=& \frac{4}{R^2}\left\{ 2(\Omega-\Phi)^2 + 4\Phi^2 + 2(\Phi+\omega\Omega)^2 + \left[2\Phi - (1+\omega)\Omega\right] \right\}.
    \end{eqnarray}
\end{subequations}
The Ricci scalar contains informations about the energy content of the spacetime. In particular, in our framework $R^{(4)}$ vanishes if the Universe is empty or if its energy content is described by an EOS parameter $\omega =1/3$. The Kretschmann scalar instead is a useful probe to use to investigate spacetime singularities. 

Constant-time slices of the spacetime provide an easier visual understanding of the geometry of the Universe. We compute the Ricci curvature of the spatial sections as
\begin{equation}\label{eq:3_curvature}
    R^{(3)} \equiv R_{ij\phantom{\ell}k}^{\phantom{ij}\ell} g^i_{\phantom{i}\ell}g^{jk} = \frac{2}{R^2}\left[ 2\Omega - U^2\left(1+2\frac{d\ln U}{d\ln\xi}\right) \right], 
\end{equation}
where $i, j, \ell, k \in \{1,2,3\}$. In Section~\ref{sec:newton} we will show how the Ricci curvature of fixed-time slices is connected to the energy under-density inside the bubble.

\section{Matching conditions}\label{sec:matchin_conditions}
For a given wall position, the system~\eqref{equations_motion_xi} is sufficient to solve the evolution of the bubble in both phases. At the wall, the order parameter $\langle \phi \rangle$ moves from the symmetric to the broken phase in a continuous way, obeying~\eqref{eq:KG}~\cite{Kurki-Suonio:1995yaf}. However, as the size of the bubble grows larger and larger, we are not interested in resolving the width of the wall, and we can replace the continuous transition of the order parameter with a discontinuous jump at the wall. The interface between the the symmetric and broken phase is then a singular surface layer described by a $\delta$-function singularity in the stress-energy tensor $T_{\mu\nu}$. 
The set of rules that connects quantities on the two sides of the layer are given by the Israel junction conditions~\cite{Israel:1966rt}.

\subsection{Junction conditions}\label{sec:junction_conditions}
In this section we give a brief summary of the Israel junction conditions, while a deeper discussion can be found in the Appendix~\ref{appendix_jc}. Let $y^\mu = (t,r, \theta, \phi)$ be the coordinate system of the global space-time, and $x^a = (\tau, \theta, \phi)$ a coordinate system on the interface of the transition (namely the surface defining the wall). Such surface is defined by a constraint $\mathcal{C}(y^\mu) = 0$, and defines a tetrad of orthogonal vectors $(n^\mu, e^\mu_\tau, e^\mu_\theta, e^\mu_\phi)$  foliating the global space-time, with $n^\mu$ the orthogonal vector to the hypersurface, and $e^\mu_a$ the tangent vectors to it. The two Israel junction conditions are~\cite{Israel:1966rt, Poisson:2009pwt}
\begin{subequations}\label{eq:JC}
    \begin{eqnarray}
        \left[h_{ab} \right]^\pm &=& 0, \\
        \left[K^{a}_{\phantom{a}b} \right]^\pm &=& - 8\pi \left(S^{a}_{\phantom{a}b} - \frac{1}{2}g^{a}_{\phantom{a}b} S \right),\label{eq:K}
    \end{eqnarray}
\end{subequations}
with
\begin{equation}
    h_{ab} \equiv g_{\mu\nu}e^{\mu}_{\phantom{\mu}a} e^{\nu}_{\phantom{\nu}b}
\end{equation}
the induced metric on the hypersurface (the first fundamental form), and
\begin{equation}
    K_{ab} \equiv e^{\mu}_{\phantom{\mu}a} e^{\nu}_{\phantom{\nu}b} \nabla_\mu n_\nu
\end{equation}
the extrinsic curvature (the second fundamental form). Moreover, we have defined the surface stress-energy tensor $S_{ab}$ as the singular part of the energy momentum tensor at the interface
\begin{equation}
    T_{\mu\nu}(y) = wu_\mu u_\nu + p g_{\mu\nu} + S_{\mu\nu} \,\delta(\mathcal{C}(y)).
\end{equation}
Combining~\eqref{eq:K} with the Gauss-Codazzi equations one finds~\cite{Berezin:1987bc} 
\begin{subequations}\label{eq:Godazzi}
    \begin{eqnarray}
        \mathcal{D}_bS^{b}_{\phantom{b}a} &=& - \left[T_{\mu\nu} e^{\mu}_{\phantom{\mu}a} n^\nu\right]^\pm \equiv -\left[T_{a}^n \right]^\pm\\
        S^{a}_{\phantom{a}b} \left\{K^{b}_{\phantom{b}a} \right\}^\pm &=& -  \left[T_{\mu\nu} n^\mu n^\nu\right]^\pm \equiv - \left[T_{n}^n \right]^\pm
    \end{eqnarray}
\end{subequations}
with $\mathcal{D}$ the covariant derivative induced by the intrinsic metric. Solving equations~\eqref{eq:JC} and~\eqref{eq:Godazzi} simultaneously is a difficult problem, and in general requires an involved numerical analysis. For the current purposes, we can apply some simplifications given the difference in scales between the size of the bubble $R_b$ and the width of the wall $W$, and expand the junction conditions at zero order in the expansion parameter $\epsilon = W/R_b \ll 1$

First, we suppose that the energy density $e$ is everywhere the same order of magnitude, including the region in the domain wall.
Combining the Friedmann equation in geometrized units $H^2\sim e$ and the requirement that the bubble remains sub-horizon $R_b H \lesssim 1$, we estimate the energy density scale of the bubble as $e \lesssim R_b^{-2}$. By dimensional analysis we estimate the surface stress energy tensor on the hypersurface as $S \sim e W \lesssim \epsilon R_W^{-1} \ll R_b^{-1}$. On the other hand, since $R_b$ is the only length scale that the intrinsic metric is sensitive to, we estimate $\mathcal{D}, K \sim R_b^{-1}$. Therefore the right hand side of equation~\eqref{eq:K} can be neglected compared to $K$. Moreover, the left hand sides of equations~\eqref{eq:Godazzi} can be estimated in this way as $\mathcal{D}S \sim KS \sim e \epsilon \ll e$. On the other hand the non-zero components of the energy momentum tensor $T\sim e$, and, as before, at zeroth order in $\epsilon$ we neglect the right hand side of equations~\eqref{eq:Godazzi}. 

The full system of junction conditions that we consider is then 
\begin{subequations}\label{eq:JC_mink}
    \begin{gather}
        \left[h_{ab}\right]^\pm = 0, \\
        \left[K_{ab}\right]^\pm = 0,\label{junc_K} \\
        \left[T_{a}^n \right]^\pm = 0, \label{eq:T_ni}\\
        \left[T_{n}^n \right]^\pm = 0.  \label{eq:T_nn}
    \end{gather}
\end{subequations}
The last two equations in~\eqref{eq:JC_mink} are analogous to the junction condition~\eqref{eq:junction_Mink} in Minkowski, while the first two equations~\eqref{eq:JC_mink}, trivially satisfied in Minkowski, are needed to quantify the discontinuity in the additional degrees of freedom in the metric $g_{\mu\nu}$. Considering the constraint $\mathcal{C}(x^\mu) = R - \xi t =0$ we find
\begin{subequations}\label{eq:e}
\begin{eqnarray}
    n^\mu &=& \gamma \left(\frac{v}{a}, \frac{1}{b}, 0,0\right)\\
    e^\mu_{\phantom{\mu}\tau} &=& \gamma \left(\frac{1}{a}, \frac{v}{b}, 0, 0\right)\\
    e^\mu_{\phantom{\mu}\theta} &=& \left(0, 0, 1, 0\right)\\
    e^\mu_{\phantom{\mu}\phi} &=& \left(0, 0, 0, 1\right)
\end{eqnarray}
\end{subequations}
with $v$ the inward fluid velocity in the constant $\xi$ frame~\eqref{def:v} and $\gamma = \left(1-v^2\right)^{-1/2}$. At $\xi =\xi_w$, the constant $\xi$ observer is at rest with the wall, so that $v(\xi_w)$ is understood as the inward component of fluid velocity in the wall rest frame. Inserting~\eqref{eq:e} and the metric~\eqref{eq:metric} in the junction conditions~\eqref{eq:JC_mink} we find (see Appendix~\ref{appendix_jc} for a detailed derivation of~\eqref{eq:junction_R} and~\eqref{eq:junction_phi})
\begin{subequations}\label{eq:JC_mink_finale}
    \begin{gather}
        \left[R\right]^\pm = 0, \label{eq:junction_R}\\
        \left[\Phi\right]^\pm = 0,\label{eq:junction_phi} \\
        \left[(1+\omega)\Omega v\gamma^2 \right]^\pm = 0, \label{eq:JC_mink_1}\\
        \left[(1+\omega)\Omega v^2\gamma^2 + \omega\Omega \right]^\pm = 0.  \label{eq:JC_mink_2}
    \end{gather}
\end{subequations}
We can understand the continuity of $R$ from the fact that $4\pi R^2$ is the proper area of a spherical shell at fixed time $t$, and two observers on the different sides of the interface must agree on the measurement of the proper area. The continuity of $\Phi$ instead derives from the fact that the mass $M$ is computed through~\eqref{eq:drM} as the integral of the energy density $\Omega$ from the origin of coordinates to $R$. A finite jump in $\Omega$ at the interface would then produce a discontinuity in the first derivative of $\Phi$, without spoiling the continuity of the function itself. 

At first order in the limit $\epsilon = W/R_b \ll 1$, the junction conditions~\eqref{eq:JC_mink_1} and~\eqref{eq:JC_mink_2} share the same functional structure of the ones in a flat Minkowski space-time~\cite{Hindmarsh:2016lnk, Hindmarsh:2019phv}. Hence, solutions~\eqref{eq:v_+} and~\eqref{eq:v_-} apply for $v(\xi_w)$ with strength parameter $\alpha_+$ 
\begin{equation}\label{eq:alpha_2}
    \alpha_+ =  \frac{\Omega_+ (1-3\omega_+) - \Omega_-(1-3\omega_-)}{3\Omega_+(1+\omega_+)} .
\end{equation}
Now, from~\eqref{eq:JC_mink_1}, we can extract the energy density outside the interface
\begin{equation}
    \Omega_+ =  \frac{(1+\omega_-)}{(1+\omega_+)} \frac{v_-\gamma_-^2}{v_+\gamma_+^2}\Omega_-.
\end{equation}
Finally, inverting~\eqref{def:v}, we get  a quadratic equation for $U_+$ in terms of the other variables. The positive root is
\begin{equation}
    U_+ = \sqrt{\frac{1-2\Phi_+}{(\Phi+\omega\Omega)^2_+ - v_+^2(\Omega-\Phi)_+^2}} (\Omega - \Phi)_+ v_+,
\end{equation}
while the negative root for $U_+$ can be neglected since it represents an inflowing solution in the non-comoving Eulerian frame.

\subsection{Asymptotic FLRW solutions}\label{sec:asymptotic_FLRW}
At the centre of the bubble $r=0$ and we expect that the Schwarzschild circumference coordinate $R$ and $\Omega$ tend to zero. Moreover, the spherical symmetry of the system imposes the vanishing of the fluid velocity $U$ at the centre of the bubble, while the physical interpretation of $M$ as a mass inside a shell of radius $R$ suggests that $M$ (and then $\Phi$) tends to zero in this limit. This motivates us to parametrize $ U = j\xi^\gamma, \; \Phi = k\xi^\alpha, \; \Omega = \ell \xi^\beta$ in a sufficiently small neighborhood of the origin $\xi = 0$, with $\alpha, \beta, \gamma, j, k, \ell$ unknown constants. The equations of motion~\eqref{equations_motion_xi} with the gauge condition $a(t,r=0) =1$ allow to restrict the solutions to a one parameter family
\begin{subequations}\label{initial_conditions}
\begin{eqnarray}
    U(\xi\rightarrow 0) &=& \frac{2}{3(1+\omega_-)}\xi,\label{eq:U_0}\\
    \Omega(\xi \rightarrow 0) &=& 3k\xi^2 \label{eq:Omega_0},\\
    \Phi(\xi \rightarrow 0) &=& k\xi^2. \label{eq:Phi_0}
\end{eqnarray}
\end{subequations}
Since $R\geq 0$, the weak (WEC) and strong (SEC) energy conditions~\cite{Carroll:2004st, Poisson:2009pwt} demand $k>0$. The value of $k$ controls the amount of curvature of the spatial sections near the origin. Using the asymptotic solution~\eqref{initial_conditions} in the expression~\eqref{eq:3_curvature}, we find
\begin{equation}\label{eq:R_3_0}
    R^{(3)}_0 \equiv R^{(3)}(\xi\rightarrow 0)  = \frac{12 \xi^2}{R^2}\left[ k-\frac{2}{9(1+\omega_-)^2} \right].
\end{equation}
This equation highlights the presence of a unique value $k_F = 2/9(1+\omega_-)^2$ for which the spatial sections around the origin are flat. For $k>k_F$ the spatial curvature is positive, while for $k<k_F$ the spatial curvature is negative. 
As the Minkowski solutions manifest a deficit in the energy density in the interior of the bubbles, following this analogy we are led to consider only the values of $k$ smaller than $k_F$, that is
\begin{equation}\label{eq:k}
    0 < k < k_F.
\end{equation}

Once the boundary conditions at the origin are fixed, the system~\eqref{equations_motion_xi} can be integrated out to the wall, where the junction conditions~\eqref{eq:JC_mink_finale} provide the matching between the two phases and the initial conditions for the solutions outside the wall. Outside the wall the system can be integrated until it is matched via a shock to a spatially flat Friedmann-Lemaître-Robertson-Walker (FLRW) Universe. The outermost part of the solution is described by the metric~\eqref{eq:metric} with 
\begin{eqnarray}
    a_F(t,r) &=& a_F \\
    b_F(t,r) &=& s(t) \\
    R_F(t,r) &=& rs(t)
\end{eqnarray}
Here, the subscript $F$ denotes quantities evaluated on a FLRW background. Then the Friedmann equations imply
\begin{gather}\label{eq:FLRW}
    U_F^2 = 2\Phi_F, \qquad \Omega_F = 3\Phi_F,
\end{gather}
and given the solution
\begin{equation}
    H=\left(\frac{\dot{s}}{s}\right) = \frac{2}{3(1+\omega)}\frac{1}{t} ,
\end{equation}
we can directly compute 
\begin{equation}\label{def:U_bg}
    U_F = \frac{2}{3(1+\omega)}\frac{\xi}{a_F}.
\end{equation}

Outer constant-$\xi$ observers live in a FLRW Universe and are described by the 4-velocity~\eqref{def:const_xi} with the additional conditions~\eqref{eq:FLRW}. The outer observers can signal inwards to the inner constant-$\xi$ observer inside the bubble about the Universe they live in; in particular they can communicate the boost that inner observers should apply in order for them to be comoving with what would be the cosmic fluid in an unperturbed FLRW Universe. 
There then exists a class of observers $U^\mu$ that are comoving with what would be cosmic fluid in a FLRW Universe defined by
\begin{equation}
    U^\mu \equiv \Lambda^\mu_{\phantom{\mu}\nu}(v_F) V_\xi^\nu
\end{equation}
where we denoted the boost transformation 
\begin{equation}
    \Lambda^\mu_{\phantom{\mu}\nu} (v_F) = \gamma_F\begin{pmatrix}
1 & -\frac{b}{a}v_F & 0 & 0\\
-\frac{a}{b}v_F & 1 & 0 & 0 \\
 0 & 0 & 1 & 0 \\
 0 & 0 & 0 & 1
\end{pmatrix}
\end{equation}
and $v_F$ the value of $v$ at $\xi$ in a FLRW Universe with equation of state parameter $\omega$
\begin{equation}\label{def:v_F}
    v_F = \frac{1+3\omega}{3(1+\omega)} \frac{\xi}{a_F}, \qquad \gamma_F = \frac{1}{\sqrt{1-v_F^2}}.
\end{equation}
After some simple algebra we find
\begin{equation}\label{def:U_mu_obs}
    U^\mu = \gamma_u \left( \frac{1}{a}, \frac{u}{b}, 0, 0\right),
\end{equation}
with
\begin{equation}
    u \equiv \frac{v - v_{F}}{1 - vv_{F}}, \qquad \gamma_u =  \frac{1}{\sqrt{1-u^2}}.
\end{equation}
the relative velocity between a constant-$\xi$ observer $V_\xi^\mu$ and what would be a constant-$\xi$ in an unperturbed FLRW Universe ${V_F}_\xi^\mu = \gamma_F(a^{-1}, b^{-1}v_F, 0, 0)$:
\begin{equation}
    v_\textnormal{rel}^2 (V_\xi,{V_F}_\xi) = u^2 .
\end{equation}
By Lorentz symmetry, this coincides with the relative fluid speed that the observers $V_\xi$ and ${V_F}_\xi$ would measure. Hence, $u$ represents the local departure from a FLRW Universe that outer constant-$\xi$ observers can communicate to the inner constant-$\xi$ observers. By construction, $u$ vanishes outside the shock, where the Universe is described by the flat FLRW solution.

\section{Properties of self-similar solutions}

\subsection{Fixed points} 
\label{sec:critical_point}
In the case of a Minkowski background, see Section~\ref{sec:mink}, the equations of motion~\eqref{eq_motion_mink} for the fluid velocity and energy density show three fixed points~\cite{Hindmarsh:2016lnk, Hindmarsh:2019phv, Cai:2018teh}. In analogy with the Minkowski case, in order to look for fixed points, it is convenient to introduce an auxiliary parameter $\tau$ such that
\begin{subequations}\label{equation_motion_tau}
    \begin{eqnarray}
    \dfrac{d\ln\xi}{d\tau} &=& \left[ U^2(\Phi+\omega\Omega)^2 - c_s^2\Gamma^2(\Omega-\Phi)^2\right],\\
    \dfrac{d\ln U}{d\tau} &=&  \left[ (\Phi+\omega\Omega)^2-2c_s^2\Gamma^2\Phi \right] \left[\Omega - \Phi\right],\\
    \dfrac{d\ln\Omega}{d \tau}  &=& \dfrac{\Omega - \Phi}{\Phi+\omega\Omega}\left[ 2\omega \frac{d\ln\xi}{d\tau} + (1+\omega)\dfrac{d \ln U}{d\tau}\right],\\
    \dfrac{d\ln\Phi}{d\tau} &=&  \frac{\left( \Omega - \Phi\right)}{\Phi}\frac{d\ln\xi}{d\tau}.
\end{eqnarray}
\end{subequations}
A fixed point is found whenever $\frac{d\xi}{d\tau} = \frac{dU}{d\tau} = 0$, as at this point the right hand side of the full system vanishes identically. We find three different solutions to the fixed point conditions. One fixed point is the trivial solution $(\xi,U,\Omega,\Phi) = (0,0,0,0)$. A second solution is
\begin{subequations}\label{critical_point}
\begin{eqnarray}
    {c_s}_- &=& \frac{U_\star}{\Gamma_\star}\frac{\Phi_\star+\omega\Omega_\star}{\Omega_\star- \Phi_\star}, \label{eq:crit_cs}\\
    U_\star &=& \frac{\Omega_\star - \Phi_\star}{\sqrt{2\Phi_\star}}, \label{eq:crit_U}
\end{eqnarray}
\end{subequations}
where the star denotes variables that satisfy the fixed point conditions  $\frac{d\xi}{d\tau} = \frac{dU}{d\tau} = 0$ and ${c_s}_-$ is the sound speed of the broken phase~\eqref{def:c_s_dist}. In light of the expression~\eqref{eq:v_expression} of the inward fluid speed for a constant-$\xi$ observer, we can replace the first condition~\eqref{eq:crit_cs} with the more convenient expression
\begin{equation}
    v_\star = {c_s}_-.
\end{equation}
The third fixed point allowed by the system~\eqref{equation_motion_tau} satisfies the fixed point conditions with $v_\star = {c_s}_-$ and $\Omega_\star = \Phi_\star$. This class of solutions is compatible either with an EOS parameter $\omega_-=-1$ or with a fluid flow $U$ that is positive around the origin, crosses $U_\star =0$ and becomes negative after the fixed point. The latter situation occurs when $k>k_s$, that is for solutions with positive spatial curvature $R^{(3)}$ around the origin.  We interpret this solution as the time reversal of the ones studied in~\cite{Musco:2012au}. However, since this fixed point does not play any role in our solutions, we will not discuss this further.

\begin{figure}
    \centering
    \includegraphics[width=\textwidth]{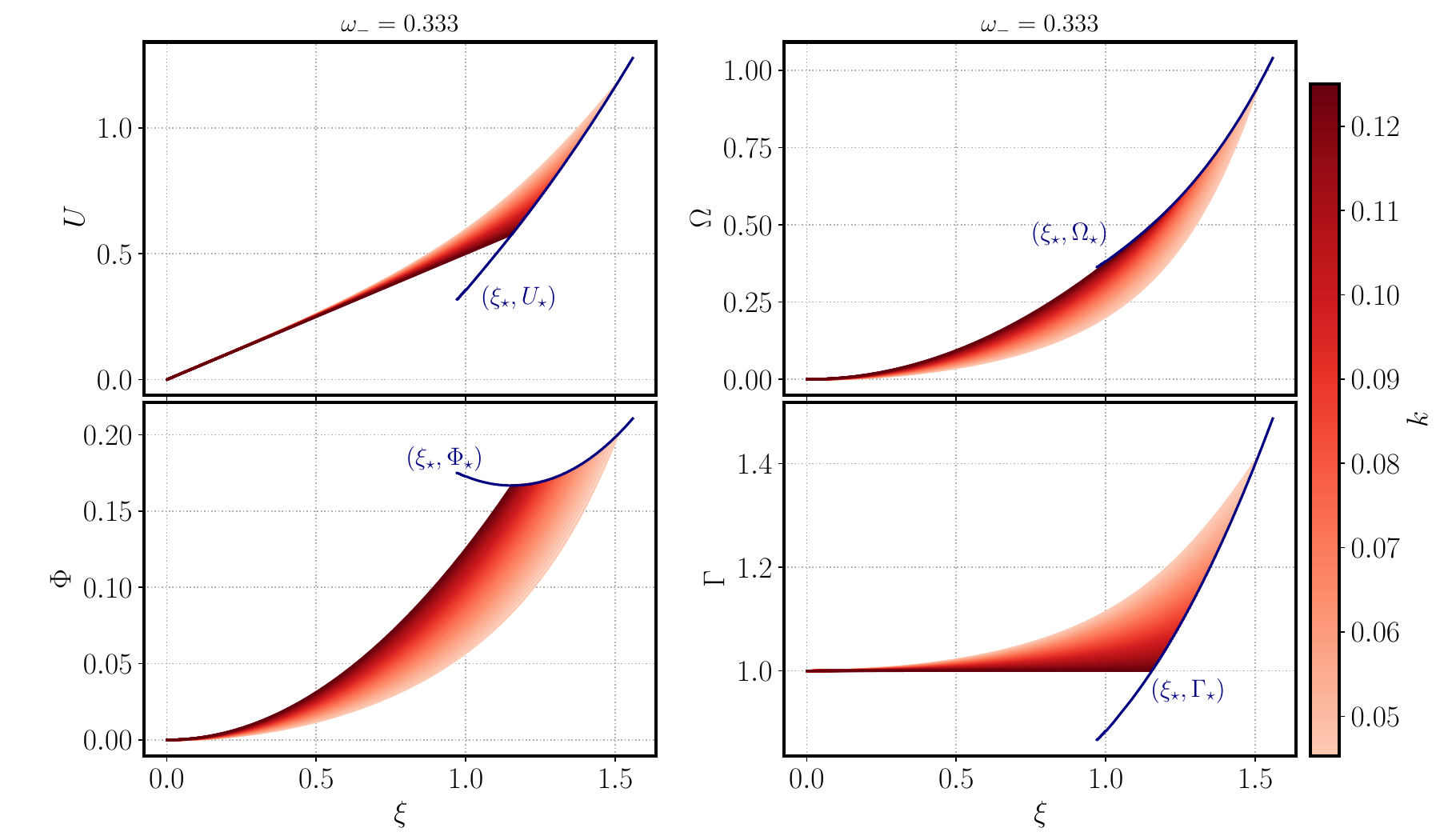}
    \caption{Profile of $U$, $\Omega$, $\Phi$ and $\Gamma$ against the self-similar variable $\xi=R/t$ in the broken phase obtained by integrating equations~\eqref{equation_motion_tau} outward from the origin with boundary conditions~\eqref{initial_conditions}. For any value of $k$, corresponding to a particular shade of red in the curves, the solutions end up on a fixed point $(\xi_\star, U_\star, \Omega_\star, \Phi_\star)$. Varying $k$, the endpoints of the solutions trace the blue lines $(\xi_\star, U_\star)$, $(\xi_\star, \Omega_\star)$, $(\xi_\star, \Phi_\star)$, $(\xi_\star, \Gamma_\star)$ in the respective projected 2-dimensional spaces.}
    \label{fig:solution_critical}
\end{figure}

\begin{figure}
    \centering
    \includegraphics[width=\textwidth]{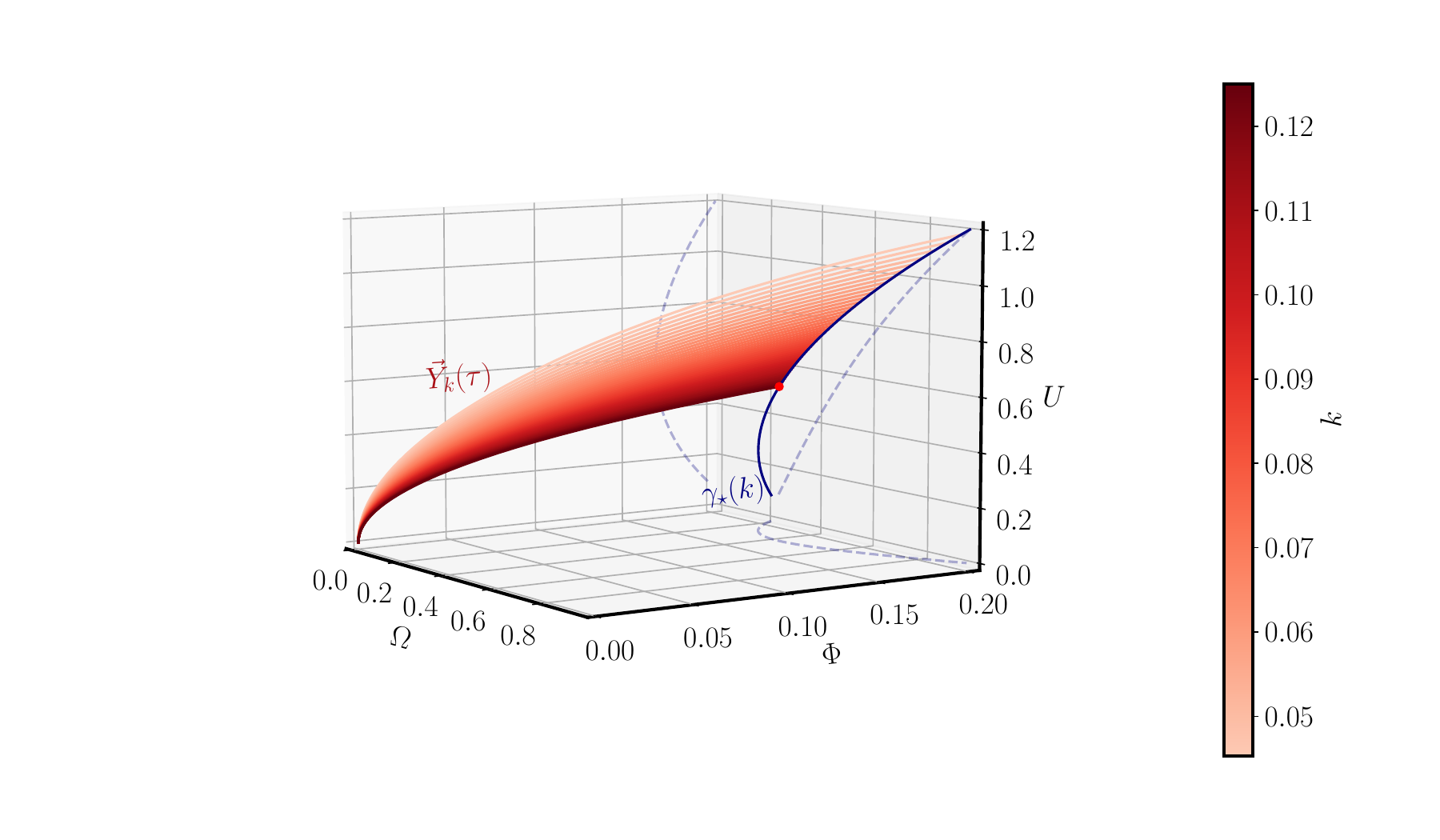}
    \caption{Trajectory of the solution $\vec{Y}_k(\tau) = (U_k(\tau), \Omega_k(\tau), \Phi_k(\tau))$ in the three-dimensional phase space in different shades of red corresponding to different initial conditions~\eqref{initial_conditions}. Varying the parameter $k$, the endpoint of $\vec{Y}_k(\tau)$ moves along the solid blue curve that represents the line of fixed points 
        of the autonomous system (\ref{equation_motion_tau}), denoted 
    $\gamma_\star(k) = (\xi_\star(k), \vec{Y}_\star(k))$. The dashed blue curves show the projection of $\gamma_\star(k)$ in the spaces $(\Omega, \Phi)$, $(U, \Phi)$, and $(U, \Omega)$.}
    \label{fig:fixed_line}
\end{figure}

Due to the $\ln\xi$ shift symmetry of the equations of motion~\eqref{equations_motion_xi}, we can constrain the study of fixed points to the phase-space $(U,\Omega,\Phi)$. Indeed the conditions $\frac{d\xi}{d\tau} = \frac{dU}{d\tau} = 0$ are met regardless of the value of $\xi$. Conditions~\eqref{critical_point} define a codimension-$2$ hypersurface in the phase space $(U, \,\Omega, \,\Phi)$, i.e. a line, that we will denote as \textit{fixed line}, or line of fixed points. $\vec{Y}_\star \equiv (U_\star,\Omega_\star,\Phi_\star)$ is then a line of fixed points for any $\xi$, and $(\xi, \vec{Y}_\star)$ is a surface of fixed point in the four-dimensional phase-space that we call a \textit{fixed surface}. Similarly, we will refer to the solution of the Einstein equation $\vec{Y}(\tau)\equiv(U, \,\Omega, \,\Phi)$ as the \textit{trajectory of the solution} in the phase space. Figure~\ref{fig:solution_critical} shows that, integrating outward from the origin with boundary conditions~\eqref{initial_conditions} for a given parameter $k$, the trajectory of the solution $\vec{Y}_k(\tau)$ always ends up on a fixed point. 
We denote the value of $\xi$ at the end point of $\vec{Y}_k(\tau)$ with $\xi_\star(k)$.  
The $\ln \xi$ shift symmetry of the equations of motion is broken by the initial conditions~\eqref{initial_conditions} with the choice $a(\xi\rightarrow 0) =1$. As mentioned above (see Sections~\ref{sec:coordinate_frame} and~\ref{sec:asymptotic_FLRW}), this choice fixes the time coordinate $t$ and the scale of $\xi=R/t$. Therefore the scale of $\xi_\star(k)$ is set by the condition $a(\xi\rightarrow 0) =1$, and the set of points $\gamma_\star(k)\equiv (\xi_\star(k), \vec{Y}_\star(k))$ is reduced to a curve in the four dimensional phase space. As one can see from Figure~\ref{fig:fixed_line} and~\ref{fig:solution_critical}, varying the parameter $k$ the end point of the trajectory of the solution $\vec{Y}$ moves along the line of fixed points $\gamma_\star(k)$. In other words, $k$ parametrizes the line of fixed points $\gamma_\star(k)$.

\subsection{Bubble solutions}
\label{sec:bubble_expansion_gr}
\subsubsection{Subsonic deflagrations}
In a deflagration the fluid speed behind the wall in the constant $\xi$ frame is always smaller than the sound speed in the broken phase, i.e.  $v_- < {c_s}_-$. Hence the trajectory of the solution never reaches the line of fixed points. At the wall, in order to match on to an inflowing solution $U_+ >0$, the negative sign in~\eqref{eq:v_+} must be chosen. The fluid outside the wall then moves slower than inside, forming a compression wave in front of the wall. In order for the bubble to expand over the symmetric phase, we need to verify that the inward fluid speed remains positive at the wall $v_+>0$. According to~\eqref{eq:v_+}, this condition puts an upper bound on the strength parameter for deflagration solutions
\begin{equation}\label{eq:upper_bound_def}
    0 <\alpha_+ < \frac{1}{3} .
\end{equation}

The compression wave must revert to a Friedmann solution within the boundary imposed by causality in order to ensure that patches of the Universe causally disconnected with the bubble remain unperturbed. Causal connection is established in the region $ v <  v_{lc} $ where $v_{lc}$ is the value of the fluid speed for constant-$\xi$ observes $v$ at the light cone. The light cone is found by looking at the region where the vector~\eqref{def:const_xi} has zero norm, i.e. $g_{\mu\nu}V^\mu_\xi V^\nu_\xi = 0$, which occurs at
\begin{equation}\label{def:light_cone}
    v_{lc} =1.
\end{equation}
The compression wave matches on to a FLRW Universe through a shock, i.e.~with a new discontinuity that must satisfy the junction conditions~\eqref{eq:JC_mink_finale} while matching to the FLRW solutions~\eqref{eq:FLRW} outside the shock, which satisfy 
\begin{equation}\label{eq:bg_shock}
    U^{sh}_+ = \sqrt{2\Phi^{sh}_+}, \qquad \Omega^{sh}_+ = 3\Phi^{sh}_+.
\end{equation}

\subsubsection{Detonation}
In a detonation, the fluid speed $v_-$ just behind the wall is always larger than the sound speed of the broken phase $ v_- > {c_s}_-$, and the fluid moves as in a FLRW Universe for constant-$\xi$ observers outside the wall, $u=0$. 
This provides a natural initial condition for the inward integration.
As in the Minkowski case, the positive sign in~\eqref{eq:v_-} must be chosen in order to ensure that the fluid moves with a supersonic velocity $ v_- > {c_s}_-$ behind the wall. The condition $v_+ >0$ in~\eqref{eq:v_+} is always satisfied with the choice of the positive sign, so that no constraint on $\alpha_+$ comes from the junction conditions.
However, according to Eq.~\eqref{eq:alpha_plus_us}, $\alpha_+>1/3$ means that the sound speed in the symmetric phase has negative square: ${c_s}_+^2<0$. Although these solutions seem to be allowed by the equations of motion~\eqref{equation_motion_tau} and the junction conditions ~\eqref{eq:v_+}, a negative sound speed squared would indicate an instability, and we therefore look for solution within the bounds~\eqref{eq:upper_bound_def}.

Equipped with the initial conditions $(U_-, \Omega_-,\Phi_-)$, we can integrate inwards towards the line of fixed points $(U_*(k), \Omega_*(k),\Phi_*(k))$, obtaining the profile of a rarefaction wave behind the wall. The asymptote of the solution at the line determines the value of the parameter $k$, and how to match to a trajectory smoothly reaching the origin.

The condition $ v_- > {c_s}_-$ implies that there is a minimum fluid velocity outside the wall given by $v_+(\alpha_+, {c_s}_-)$; this is the Chapman-Jouguet speed $v_\textnormal{CJ}$. As explained in Section~\ref{sec:eos} and after equation~\eqref{def:eos}, in our model the sound speed in the broken phase is fixed at ${c_s}_- = 1/\sqrt{3}$, for which, from equation~\eqref{eq:v_+}, one finds
\begin{equation}
    v_\textnormal{CJ} = \frac{1}{\sqrt{3}}\left( \frac{1+\sqrt{\alpha_+ + 3\alpha_+^2}}{1+\alpha_+}\right).
\end{equation}
Hence, in a detonation the fluid velocity in the constant $\xi$ frame just outside the wall $v_+$ is bounded from above and below by
\begin{equation}\label{eq:bound_v_+}
    v_\textnormal{CJ} \leq v_+ < 1 .
\end{equation}

\subsubsection{Supersonic deflagrations}
Supersonic deflagration (or hybrid) solutions are found between the two aforementioned cases when the fluid exit speed in the constant $\xi$ frame is $v_- = {c_s}_-$, while the fluid is at rest with the FLRW cosmic fluid for constant-$\xi$ observers, i.e. $u=0$, outside the bubble.
The outer FLRW Universe provides a natural initial condition for the inward integration from the shock. The fluid speed ahead of the shock for constant-$\xi$ observers is bounded by
\begin{equation}\label{eq:v_sh_boundary}
    {c_s}_- < v_+^{sh} < v_\textnormal{CJ}.
\end{equation}
In the region $ v_+^{sh} > v_\textnormal{CJ}$ the solution is a detonation, while in the region $v_+^{sh}<{c_s}_-$ the solution is a deflagration. The junction conditions are the same for subsonic and supersonic deflagrations: at the shock
\begin{equation}\label{JC_shock}
    v_-^{sh} =\frac{\omega_+}{v_+^{sh}}, \qquad \Omega_-^{sh} = {\gamma_+^{sh}}^2 ({v_+^{sh}}^2 -\omega_+^2)\frac{\Omega_+^{sh}}{\omega_+},
\end{equation}
and we integrate inward along the compression wave until the wall, where the junction condition~\eqref{eq:v_-} has to be taken with the negative sign in order to ensure that $U_+ >0$.  As in a subsonic deflagration, the bounds~\eqref{eq:upper_bound_def} apply. The junction conditions at the wall provide the initial conditions $(U_-,\Omega_-, \Phi_-)$ for a second inward integration along the rarefaction wave towards the line of fixed points $\gamma_\star(k)$. The intersection between the trajectory of the solution and the fixed line determines the value of the parameter $k$ and the matching to a trajectory that smoothly reaches the origin, as done for the detonation case. Integrating outward from the shock, the we stop the solutions at the light cone where $v =1$.

\section{Numerical methods}\label{sec:numerical_results}
In our model, the EOS inside the bubble is fixed with  $\omega_{-} = 1/3$. 
We wish to find solutions as a function of the EOS parameter in the symmetric phase $\omega_+$, which determines the strength of the transition at the wall~\eqref{eq:alpha_plus_us}, and the wall position $\xi_w$ that discriminates between the three possible cases: subsonic deflagration, detonation, and supersonic deflagration. In order to find the solutions, we must determine the parameters $k$ and $a_F$, which characterize the asymptotic solutions at the origin and at the light cone respectively, and $\xi_{sh}$, which determines the position of the shock in a deflagration. 
In this section we provide some numerical methods that we used to find our solutions and comment on the results.

In all the cases of study of this paper, we performed the integration of the system~\eqref{equations_motion_xi} using the fourth-order Runge-Kutta scheme encoded in the Python SciPy solver \textit{integrate.solve\_ivp}. We adopted different shooting algorithms to obtain the optimal values of the parameters $k$, $a_F$ and $\xi_{sh}$ using the bisection routine carried out by the SciPy method \textit{optimize.bisect}.

In all the discussed cases, the initial displacement in $\xi$ from the origin that we used for numerical integration is $10^{-3}$. To maximize the precision of the solutions we use the minimum relative (\texttt{rtol = 2.3e-14}) and absolute (\texttt{atol =1E-18}) tolerance levels allowed by the solver \textit{integrate.solve\_ivp}. We will explain in the following paragraphs that the position of the shock (wall) in a subsonic (supersonic) deflagration as well as the position of the fixed line in detonation and hybrid solutions are dynamically found when the solver \textit{integrate.solve\_ivp} meets certain requirements. The precision with which we can fix the interfaces and the fixed line is controlled by the sampling spacing $\Delta\xi$ in the $\xi$ coordinate. A smaller sampling spacing provides a better resolution, at the price of a lower efficiency of the code. Depending on the case, we aim to find the best compromise, keeping as an upper bound $\Delta\xi <10^{-4}$.

\subsection{Subsonic deflagration}
Integrating outward from the origin with fixed $\omega_+$ and $\xi_w$, we shoot with the parameter $k$ to determine the shock position $\xi_{sh}$. The parameter $a_F$ can be read afterwards from the junction conditions at the shock, but it does not play any role in the shooting procedure for deflagrations. Keeping in mind that the solution must match on to a FLRW Universe~\eqref{eq:FLRW} outside the shock, the shooting algorithm for optimal values of $k$ and $\xi_{sh}$ proceeds as follow:
\begin{enumerate}
    \item for a given value of $k$ in the range~\eqref{eq:k} we integrate forward from the origin up to the wall.
    \item The junction conditions~\eqref{eq:JC_mink_finale} at the wall give the boundary conditions for the trajectory of the solutions $\vec{Y}$ just outside the wall. We integrate forward until the first shock condition~\eqref{eq:T_nn} $\left[T^{n}_{\phantom{n}n}\right]^\pm_{sh}= 0$ is met, where 
    \begin{equation}
        \left[T^{n}_{\phantom{n}n}\right]^\pm_{sh} \equiv T^{n}_{\phantom{n}n}\left(\sqrt{2\Phi^-_{sh}}, 3\Phi^-_{sh}, \Phi^-_{sh}\right) - T^{n}_{\phantom{n}n}\left(U^-_{sh},\Omega^-_{sh},\Phi^-_{sh}\right), 
    \end{equation}
    and we used the fact that the junction conditions~\eqref{eq:junction_Mink} imply $\Phi^+_{sh} = \Phi^-_{sh}$ and outside the shock the relations~\eqref{eq:FLRW} hold.

    \item  At this same point we evaluate the jump $\left[T^{n}_{\phantom{n}0}\right]^\pm_{sh}$~\eqref{eq:T_ni}, which turns out to be approximately a linear function of $k$ with a unique zero.

    \item We use a bisection algorithm to find the unique value of $k$ for which $\left[T^{n}_{\phantom{n}0}\right]^\pm_{sh} = 0$ at the same value of $\xi$ where $\left[T^{n}_{\phantom{n}n}\right]^\pm_{sh} = 0$, knowing that $k$ is bracketed by~\eqref{eq:k}. The value of $\xi$ at which this condition is realized defines the shock position $\xi_{sh}$. 

    \item The trajectory of solutions $\vec{Y}^+_{sh}$ just outside the shock given by the solution of the junction conditions~\eqref{eq:junction_Mink} determines the value of $a_F$ through~\eqref{def:U_bg}.
\end{enumerate}
All the unknown parameters, $k$, $\xi_{sh}$ and $a_F$, are thus set.

\subsection{Detonation}
In Section~\ref{sec:critical_point} we have shown that all the solutions starting from the origin with initial conditions~\eqref{initial_conditions} end on the line of fixed points $\gamma_\star(k)$, and $k$ is a continuous parameter of the curve. The result of this analysis is shown in Figure~\ref{fig:fixed_line}. The strategy is to split the shooting algorithm in two sections. First we integrate inward from the wall aiming to reach the fixed line $\gamma_\star(k)$, and then integrate outward from the origin aiming for the same point on the fixed line. As there is no shock in a detonation, the only two parameters to set are $k$ and $a_F$. In details, the algorithm works as follow:
\begin{enumerate}
    \item For a fixed value of $\xi_w$ we select an arbitrary value of $a_F$ in the range~\eqref{eq:bound_v_+} that defines the trajectory of solutions $\Vec{Y}$ outside the wall . The junction conditions~\eqref{eq:junction_Mink} give the boundary conditions for the trajectory $\vec{Y}$ just behind the wall.

    \item We integrate inward from the wall until the condition  $v ={c_s}_-$ is met. The end point of $\vec{Y}$ is now on a fixed point\footnote{Even though we just require one condition, $v ={c_s}_-$, the second condition~\eqref{eq:crit_U} is automatically verified by the numerical solutions. This ensures us that the end point of $\vec{Y}$ lies on a fixed point}.

    \item By direct inspection, we find that the value of $\xi$ at the end point of the trajectory of the solution, that we will denote as $\xi_\textnormal{end}$, is approximately a linear function of $a_F$, while the values of the other three variables $\vec{Y}(\xi_\textnormal{end})$ at the end point are approximately independent of $a_F$.

    \item There is a unique value of $a_F$ such that $\xi_\textnormal{end} =\xi_\star$, that is such that $(\xi_\textnormal{end}, \vec{Y}(\xi_\textnormal{end}))$ reaches the fixed point line $\gamma_\star(k)$. We find the optimal parameter $a_F$ with a bisection method, using the fact that $a_F$ is bracketed by~\eqref{eq:bound_v_+}.

    \item The intersection between the trajectory of the solution and the fixed point line identifies a point $(\xi^\textnormal{end}_{\star}, \vec{Y}^\textnormal{end}_{\star})$ which allows us to extract the parameter $k$ by inverting the parametrization of the curve $k = \gamma^{-1}_\star (\xi^\textnormal{end}_{\star}, \vec{Y}^\textnormal{end}_{\star})$.
\end{enumerate}
The unknown parameters $k$ and $a_F$ are thus determined. 

\subsection{Supersonic deflagrations}\label{sec:nm_hybrids}
It was mentioned in Section~\ref{sec:bubble_expansion_gr} that, differently from the subsonic deflagration case, the asymptotic FLRW solution outside the bubble provides a natural choice of initial conditions for the inward integration from the shock. The most convenient choice of input parameters is then $\omega_+$ and shock position $\xi_{sh}$, while through the shooting algorithm we aim to determine the metric coefficient $a_F$, the wall position $\xi_{w}$ and the critical density parameter at the origin $k$. As in the detonation case, we use the analysis of fixed points in Figure~\ref{fig:fixed_line} to split the shooting algorithm in two different sections, the first for $a_F$ and $\xi_w$, and the second for $k$. In the following we explain the algorithm for supersonic deflagrations step by step:
\begin{enumerate}
    \item With fixed $\omega_+$ and $\xi_{sh}$, we choose a value of $a_F$ that respects the bounds~\eqref{eq:v_sh_boundary}. This parameter uniquely defines the trajectory of solutions $\vec{Y}$ outside the shock.

    \item The junction conditions~\eqref{JC_shock} at the shock provide the initial condition for the inward integration.

    \item Integrate inward until the wall position $\xi_w$ defined by $v_- ={c_s}_-$. According to~\eqref{eq:v_-}, this condition is realized when 
    \begin{equation}
        \left(1+\alpha_+\right)v_+ + \frac{1-3\alpha_+}{3v_+} = \frac{2}{\sqrt{3}}.
    \end{equation}
    At each step of the integration we verify whether the above condition is met. The point in $\xi$ where this happen defines the wall position, and there we stop the integration.

    \item The junction conditions~\eqref{eq:JC_mink_finale} at the wall ensure that $v_- = {c_s}_-$ and provide the initial condition for a second inward integration toward the fixed line. However, we discussed in Section~\ref{sec:critical_point} that $v = {c_s}_-$ is one of the two conditions that define a fixed point. To leave the fixed point we introduce a small displacement $v_- = {c_s}_- +\delta$ with $\delta = 10^{-7}$. This provides safe initial conditions for the inward integration.

    \item Integrate inward from the wall toward the fixed line until the condition $v = {c_s}_-$ is met. As in the detonation case, we verify that the end point of the trajectory of the solution $\vec{Y}$ is a fixed point. From here, the algorithm follow the same strategy of last three bullets drawn for detonation solutions.
\end{enumerate}
After this shooting routine, all the necessary parameters are fixed.

\section{Results and discussion}
\begin{figure}
    \centering
    \includegraphics[width=\textwidth]{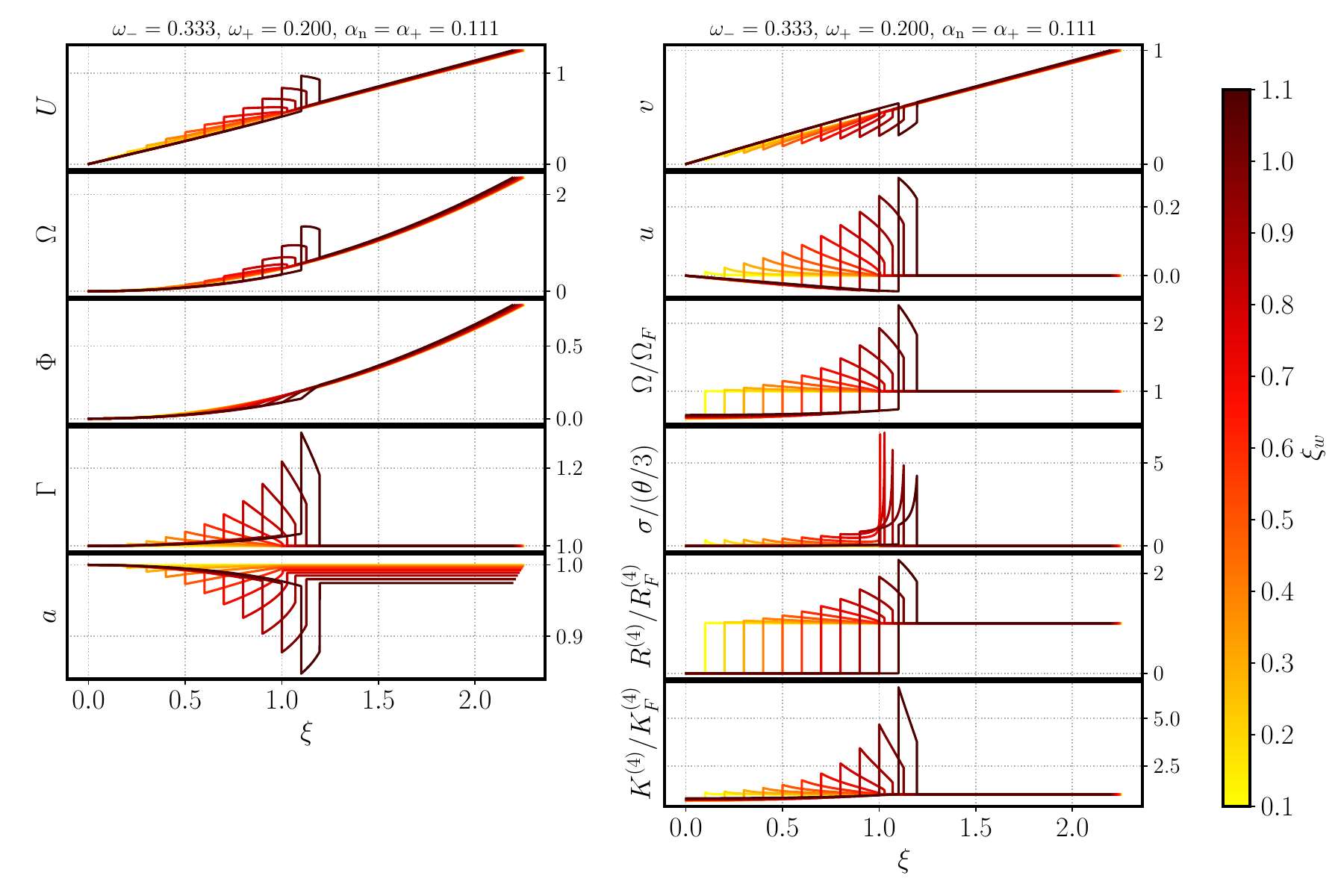}
    \caption{Deflagration solutions against self-similar variable $\xi$ for fixed EOS parameter in the broken ($\omega_{-} =1/3$) and symmetric phase ($\omega_+ = 0.2$) and different values of the wall positions $\xi_w$. On the left quantities used in the numerical integration $U, \Omega, \Phi, \Gamma, a$; on the right fluid speed $v, u$, energy $\Omega$ normalized to its value in FLRW $\Omega_F$ and shear $\sigma$ normalized to the local expansion $\theta/3$ and curvature related quantities $R^{(4)}$ and $K^{(4)}$ normalized to their respective values in the outer flat FLRW Universe $R^{(4)}_F$ and $K^{(4)}_F$.}
\label{fig:deflagrations_vw}
\end{figure}

We present in this section our numerical solutions for expanding bubbles and comment on the results. The plots in Figures~\ref{fig:deflagrations_vw} to~\ref{fig:hybrids_alpha} are organized as follow: on the left we display quantities that are directly obtained from the integration of the equations of motions~\eqref{equations_motion_xi}, $U$, the radial component of the fluid four-velocity for an Eulerian observer, $\Omega$, the energy on a shell of radius $R$, $\Phi$, the gravitational potential at $R$, $\Gamma$, the generalized Lorentz factor, and $a$, the time dilation with respect to the origin. On the right we display quantities that are derived from the previous ones. On top we put quantities related to the fluid: the fluid speed for constant-$\xi$ observers $v$ and the relative fluid speed $u$ as measured by constant-$\xi$ observers $V_\xi$ and ${V_F}_\xi$, the ratio $\Omega/\Omega_F$ between the energy on a shell of radius $R$ and its value in a FLRW Universe with EOS parameter $\omega_+$, and the shear scalar $\sigma$ normalized to the local expansion rate $(\theta/3)$. At the bottom we put quantities related to the curvature of the spacetime: the Ricci curvature scalar  $R^{(4)}$ and the Kretchmann scalar $K^{(4)}$ normalized to the values that would have in the outer FLRW Universe, $R^{(4)}$ and $K^{(4)}$ respectively. We studied three different expansion modes, subsonic deflagration, detonation, and supersonic deflagration. For each mode we perform two different analysis corresponding to two different figures for each case; in one we fix the EOS inside and outside the bubble and plot a set of solutions with different wall position $\xi_w$, while in the other we fix the wall position (shock position for hybrid solutions) and plot a set of solutions for different values of the EOS parameter $\omega_+$.

\subsection{Subsonic deflagrations}
\begin{figure}
    \centering
    \includegraphics[width=\textwidth]{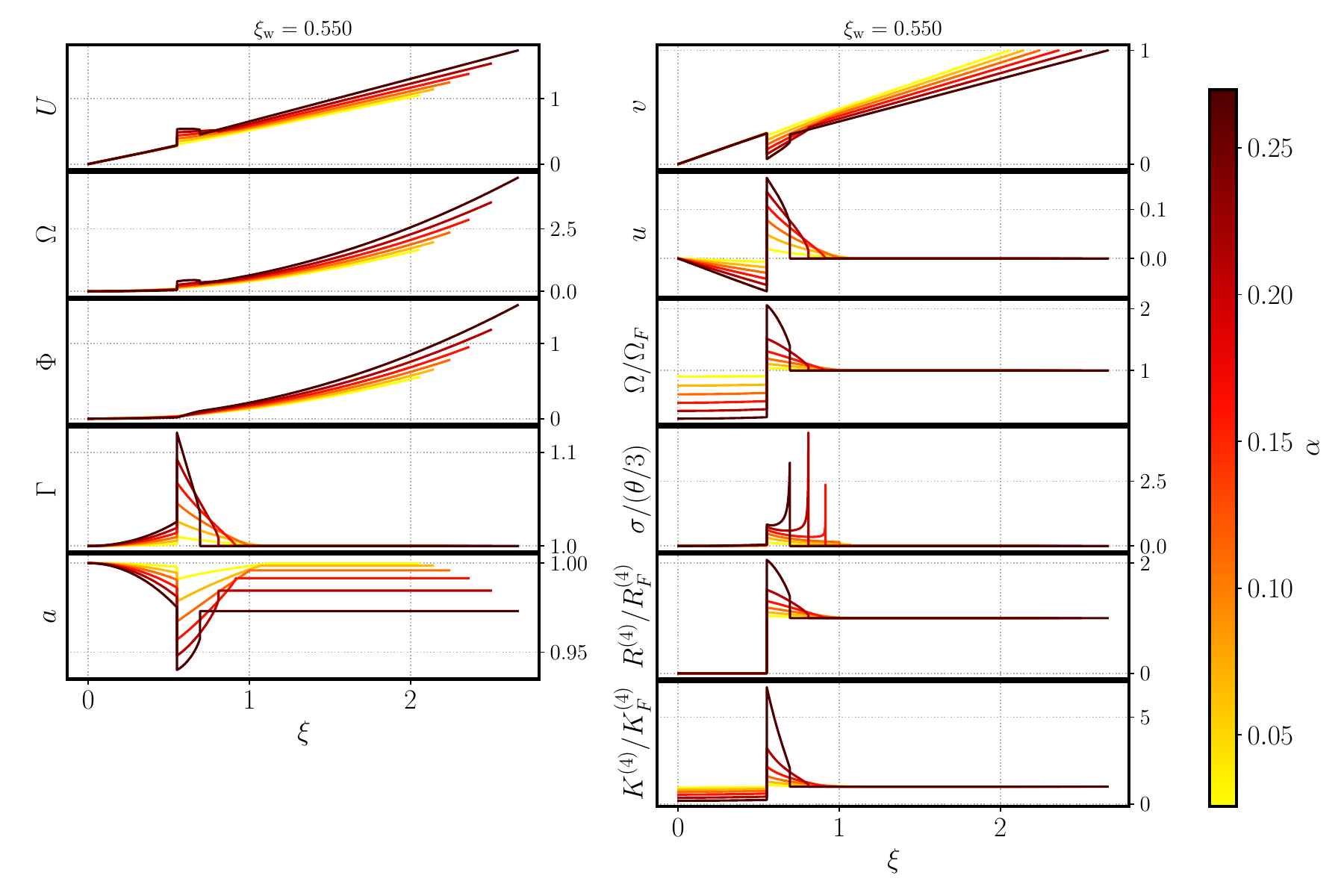}
    \caption{Deflagration solutions against self-similar variable $\xi$ for fixed wall position $\xi_w$ and varying equation of state parameter of the symmetric phase $\omega_+$. 
    See the caption to Fig.~\ref{fig:deflagrations_vw} and the text for a description of the quantities plotted.}
\label{fig:deflagration_a}
\end{figure}

We show numerical results for deflagration solutions with fixed strength $\alpha_+$ and different values of $\xi_w$ in Figure~\ref{fig:deflagrations_vw}, while deflagrations with fixed $\xi_w$ and for different values of the strength parameter at the wall $\alpha_+$ in Figure~\ref{fig:deflagration_a}.

The fluid speed $u$ contains information about the fluid velocity  as seen by constant-$\xi$ observers. Constant-$\xi$ observers outside the bubble signal to the constant-$\xi$ observers inside the bubble about the Universe they live in; inner constant-$\xi$ observers then measure a non-zero relative velocity $u$ with respect to a hypothetical FLRW constant-$\xi$ observer at the same position, which defines a rest frame for the cosmic FLRW fluid. Inspecting Figure~\ref{fig:deflagrations_vw} (top right) we see that inner constant-$\xi$ observers would measure a rarefaction wave behind the wall with respect to the cosmic FLRW fluid. Ahead of the wall, constant-$\xi$ observers see a compression wave that ends at the shock. Outside the shock, constant-$\xi$ observers measure zero relative velocity $u$, by definition, and they agree that they live in a flat FLRW Universe. In deflagration solutions in Minkowski spacetime (Figure~\ref{fig:mink}), the fluid is completely at rest with the Universe frame behind the wall. 

Information about curvature can be extracted by looking at constant-$r$ observers; these observers are comoving with the fluid and see an anisotropic expansion rate inside the bubble, as shown by the departure of the shear scalar $\sigma$ from zero in the interior of our solutions. This means that the Universe inside the bubble, is not isotropic, except the origin, and therefore cannot be reduced to any FLRW Universe. The sharp peaks in $\sigma/(\theta/3)$ at the shock come from the fact that the divisor, the local average expansion rate, becomes very small. We verify with the Kretschmann scalar $K^{(4)}$ that this does not lead to a curvature singularity. The energy ratio profile $\Omega/\Omega_F$ that constant-$r$ observers measure can be compared to the enthalpy density profile of the Minkowski spacetime solutions; in both cases, constant-$r$ observers measure a deficit of energy respect to a FLRW Universe with EOS parameter $\omega_+$ inside the wall, and an excess of energy in the compression wave. A more quantitative comparison will be performed later in Figure~\ref{fig:curvature}.

\subsection{Detonations}

\begin{figure}
    \centering
    \includegraphics[width=\textwidth]{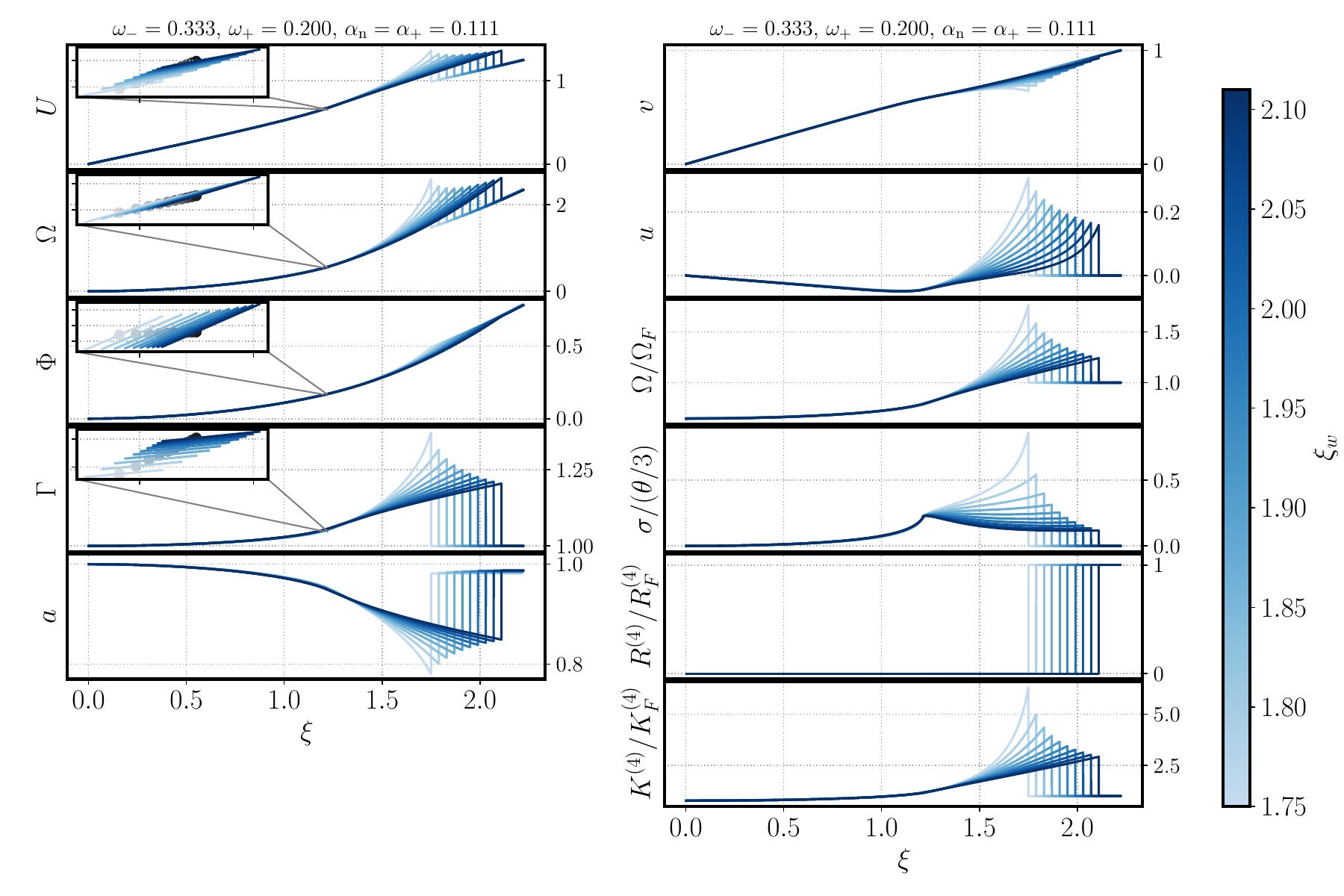}
    \caption{Detonation solutions against self-similar variable $\xi$ for fixed EOS parameter in the broken ($\omega_{-} =1/3$) and symmetric phase ($\omega_+ = 0.2$) and different values of the wall positions $\xi_w$. See the caption to Fig.~\ref{fig:deflagrations_vw} and the text for a description of the quantities plotted.
    }
\label{fig:detonation_vw}
\end{figure}

\begin{figure}
    \centering
    \includegraphics[width=\textwidth]{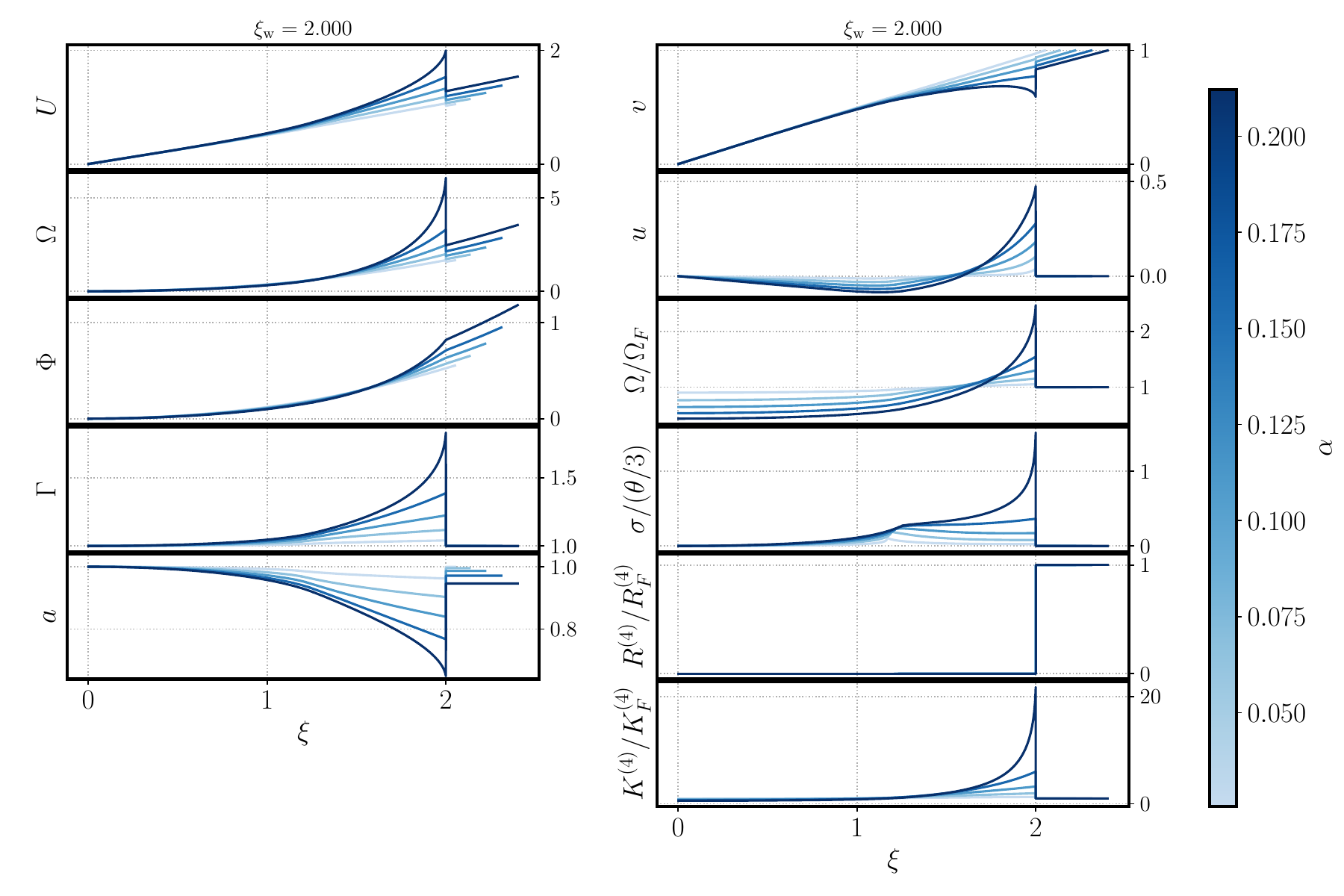}
    \caption{Detonation solutions against self-similar variable $\xi$ for fixed wall position $\xi_w$ and varying equation of state parameter of the symmetric phase $\omega_+$. 
    See the caption to Fig.~\ref{fig:deflagrations_vw} and the text for a description of the quantities plotted.}
\label{fig:detonation_alpha}
\end{figure}

In Figure~\ref{fig:detonation_vw} we display detonation solutions for different values of the wall position $\xi_w$, with fixed $\alpha_+$, and in Figure~\ref{fig:detonation_alpha} we show detonation solutions for different values of $\omega_+$ (or $\alpha_+$ equivalently) with fixed $\xi_w$.

The insets in the first four plots of Figure~\ref{fig:detonation_vw} zoom in to the region where the solutions cross the fixed line, showing the small differences in the solutions from the centre to the fixed point. 
The solutions and their derivatives are continuous there.
Detonation solutions meet the FLRW Universe just outside the wall, so that constant-$\xi$ observers do not measure a compression wave ahead of the wall. Inside the wall, $u$ differs from zero, and inner constant-$\xi$ observers measure a departure from the FLRW universe signalled in by outer constant-$\xi$ observers. While in the Minkowski spacetime solutions rarefaction waves in detonations extend inward only to the fixed point at the sound speed, in the gravitating solutions the rarefaction wave behind the wall extends to the centre of the bubble. 
We notice further that the profile of $u$ always intersects the line $u=0$ between the fixed point an the wall. This means that there exists a constant-$\xi$ observer inside the bubble that sees the fluid moving inward with the same speed that an FLRW constant-$\xi$ observer would measure. Although this observer measures a vanishing local departure from FLRW in the fluid speed, the observer can still realize that the local Universe is not FLRW by looking at the Kretschmann scalar $K^{(4)}$.
The fact that $K^{(4)}$ differs from the value that it has outside the wall means that the spacetime inside the wall is geometrically different from the spacetime outside the wall.  

The statement that the the inner and outer sides of the bubble have different geometries is also supported by the fact that constant-$r$ observers measure an anisotropic expansion $\sigma$ inside the wall. As mentioned in the deflagration case, this points to the fact that the inner side of the bubble is not a FLRW Universe except in the origin. We further notice the the position of the fixed point in the four dimensional phase space $(\xi, U, \Omega,\Phi)$ mostly depends of the EOS parameter $\omega_+$, while it is almost independent on $\xi_w$.

\subsection{Supersonic deflagrations}
\begin{figure}
    \centering
    \includegraphics[width=\textwidth]{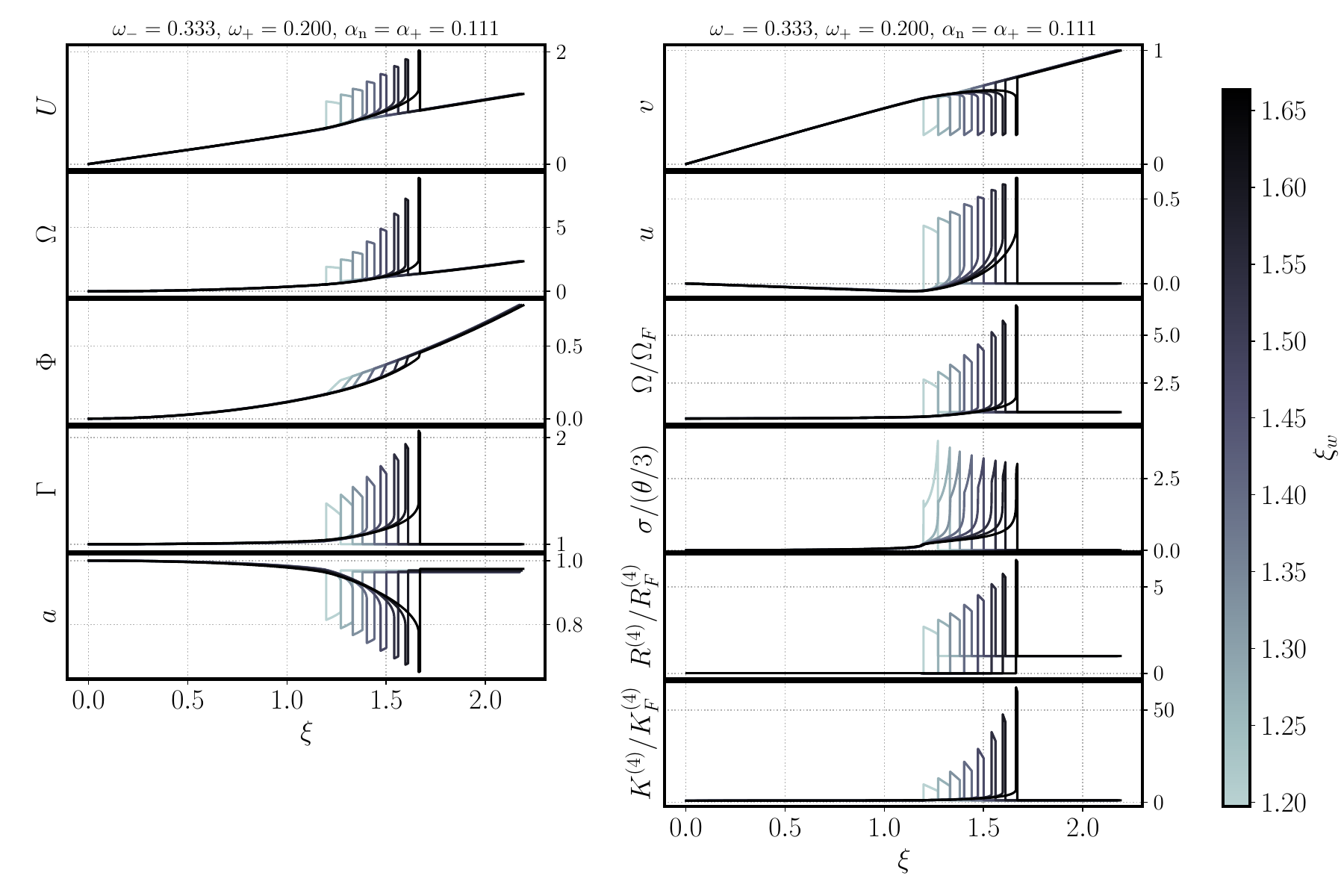}
    \caption{Supersonic deflagration solutions against self-similar variable $\xi$ for fixed EOS parameter in the broken ($\omega_{-} =1/3$) and symmetric phase ($\omega_+ = 0.2$) and different values of the wall positions $\xi_w$.   See the caption to Fig.~\ref{fig:deflagrations_vw} and the text for a description of the quantities plotted.}
\label{fig:hybrids}
\end{figure}

\begin{figure}
    \centering
    \includegraphics[width=\textwidth]{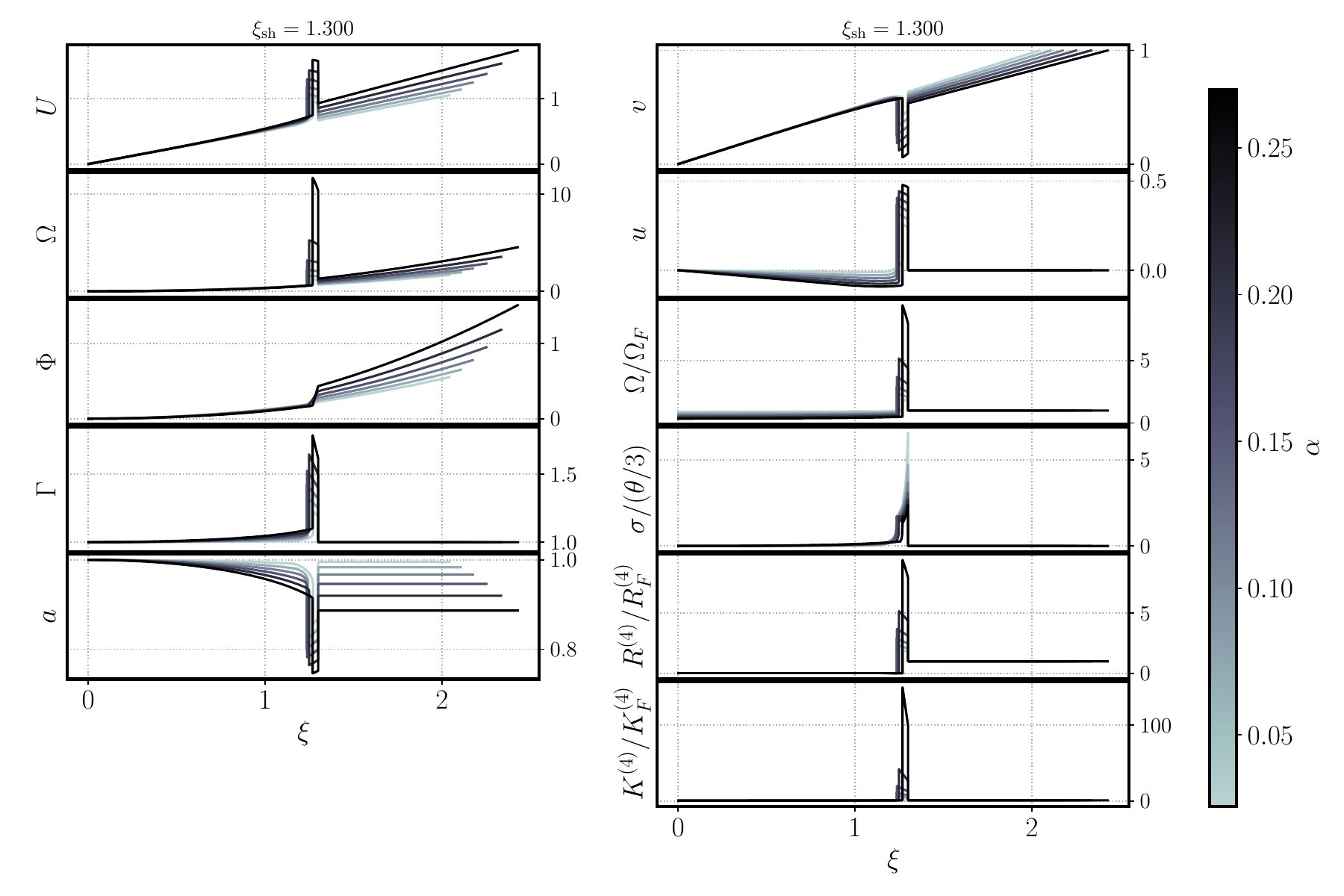}
    \caption{Supersonic deflagration solutions against self-similar variable $\xi$ for fixed shock position $\xi_{sh}$ and varying equation of state parameter of the symmetric phase $\omega_+$. 
    See the caption to Fig.~\ref{fig:deflagrations_vw} and the text for a description of the quantities plotted.}
\label{fig:hybrids_alpha}
\end{figure}

In Figure~\ref{fig:hybrids} we show some examples of hybrid solutions for different values of the shock position with fixed transition strength parameter $\alpha_+$, while in Figure~\ref{fig:hybrids_alpha} we display some example solutions with fixed shock position but different equation of state parameter in the symmetric phase $\omega_+$. 

In supersonic deflagrations the discontinuities at the wall and the shock are the strongest. Constant-$\xi$ observers measure a compression wave ahead of the wall as in the deflagration case, and a rarefaction wave behind the wall as in detonation case; the rarefaction wave passes through the fixed line and extends to the origin.

\subsection{Newtonian estimation of the curvature of spatial sections near the origin}\label{sec:newton}
In Section~\ref{sec:asymptotic_FLRW} we introduced the curvature of spatial sections near the origin~\eqref{eq:R_3_0} and its relation with the parameter $k$. In this section we want to provide a quasi-Newtonian calculation of the amount of spatial Ricci curvature that we would expect from the Minkowski spacetime solutions.

We consider linear scalar perturbations around a flat FLRW Universe~\cite{Durrer:2004fx} in the conformal Newtonian gauge~\cite{Mukhanov:1990me}
\begin{equation}
    ds^2 = s^2(\eta) \left[- (1+2\Phi_B)d\eta^2 + (1-2\Psi_B)\eta_{ij}dx^i dx^j \right],
\end{equation}
where $\eta$ is the conformal time, and the Bardeen potentials $\Phi_B$ and $\Psi_B$ are two gauge invariant scalar perturbations. The curvature of spatial sections can be then computed at fist order in scalar perturbations as
\begin{equation}
    R^{(3)} = \frac{4}{s^2}\nabla^2 \Psi_B.
\end{equation}
On the other hand, neglecting for the sake of simplicity the perturbations in the radial momentum density $T_{0i}$, the linearised Einstein equations~\cite{Durrer:2004fx, Mukhanov:1990me} reduce to the Poisson equation for the Bardeen potential
\begin{equation}
    \nabla^2\Psi_B \simeq 4\pi s^2 e \left(\frac{\delta e}{e} \right) \simeq \frac{3}{2} s^2 H^2 \delta
\end{equation}
where in the last step we used the Friedman equation $H^2 = 8\pi e/3$, $\delta e$ is the perturbation in the energy density of the fluid and $\delta = \delta e/e$. Combining the above expressions, we give the quasi-Newtonian approximation of the spatial curvature induced by the expanding bubble as
\begin{equation} \label{eq:R_3_Newton}
    R^{(3)}_M \simeq 6 H^2 \delta
\end{equation}
where $\delta$ is the energy density contrast in the Minkowski solutions, which we obtain by modifying the code snipped provided in Ref.~\cite{Giese:2020znk}. Given the barotropic EOS $p=\omega e$, we can write $\delta = w/w_1 - 1$, with $w_1$ the enthalpy outside the shock. 

In comparing the gravitating and Minkowski solutions, we run into the problem of comparing the coordinate ratios $\xi$: the gravitating solutions extend to $\xi >1$, which would be outside the light cone in Minkowski space.  We choose instead to compare by the ratio of $xi$ to its value on the light cone, $\xi_{lc}$.
In the Minkowski spacetime solutions $\xi_{lc} = 1$ always, while in the gravitating solutions the light cone position is defined by the condition $v_{lc} =1$~\eqref{def:light_cone}.

\begin{figure}
    \centering
    \includegraphics[width=\textwidth]{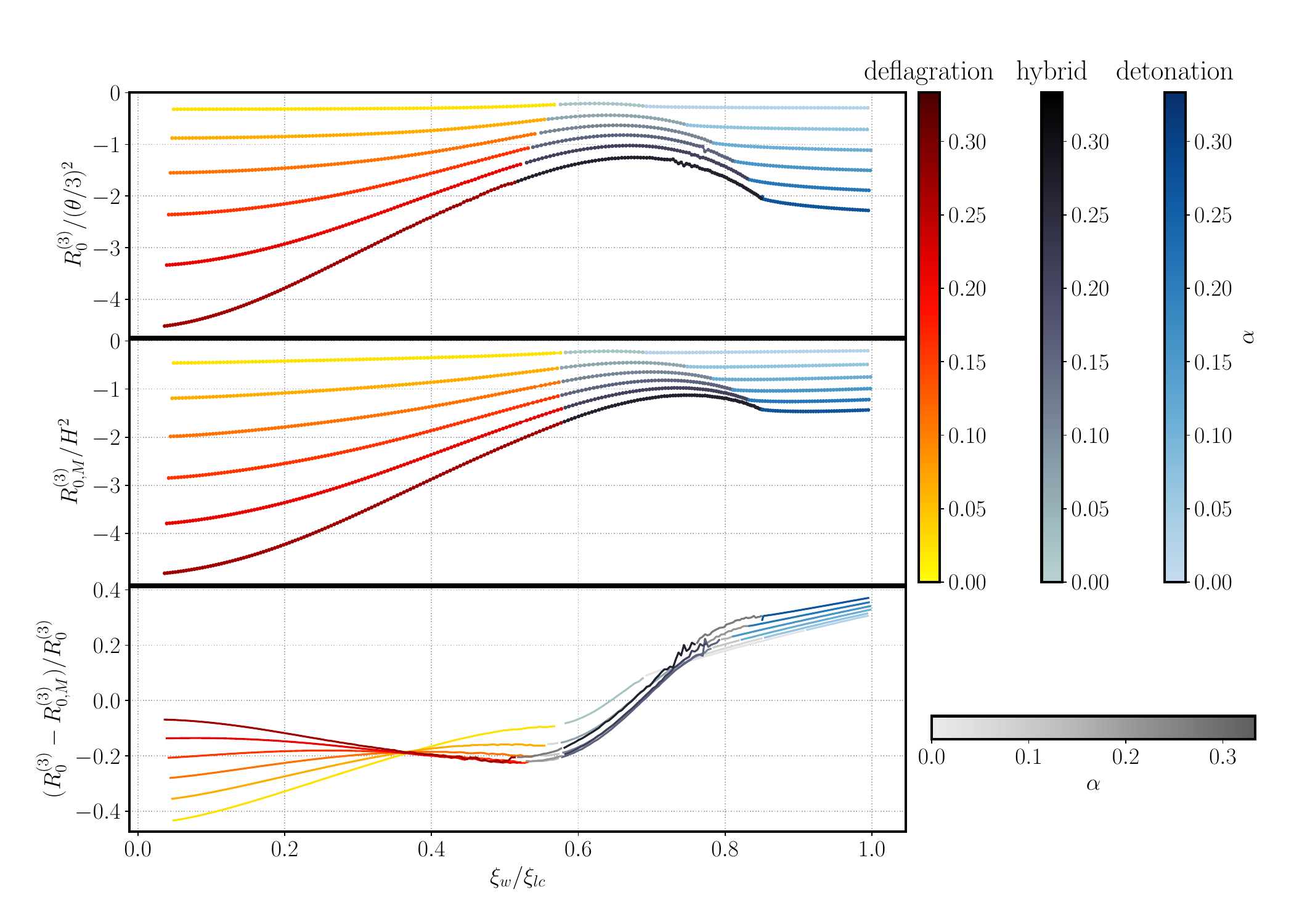}
    \caption{Ricci curvature of spatial sections evaluated at the origin, normalized by the local expansion scalar $\theta/3$, in the gravitating solutions (upper panel) and quasi-Newtionian estimate inferred by the Minkowski spacetime solutions (central panel), against the fractional distance of the wall from the light cone $\xi_w/\xi_{lc}$; in the bottom panel relative difference between the two cases above. The light cone in the Minkowski solutions is always $\xi_{lc}=1$. In a FLRW background the local expansion $\theta/3$ coincides with the global 
    Hubble constant $H$. Some of the gravitating solutions in the hybrid regime belong to the deflagration regime in the Minkowski case for the same value of $\xi_w/\xi_{lc}$. For these solutions we use a grey colormap in the bottom plot.}
    \label{fig:curvature}
\end{figure}

In Figure~\ref{fig:curvature} we plot the Ricci curvature of spatial sections near the origin $R^{(3)}(\xi\rightarrow 0)$ against $\xi_w/\xi_{lc}$ and compare it with the quasi-Newtonian estimate that we obtain from the solutions in Minkowski spacetime. 

The two cases show an overall similar dependence on the wall position $\xi/\xi_{lc}$, with a maximum in the hybrid regime. The amplitudes of the spatial Ricci curvature at the origin are around the same order of magnitude for the gravitating and quasi-Newtonian approximation cases, but gravitating deflagrations have relatively less negative curvature, and detonations more. 

Gravitating solutions are more subject to numerical errors, because the system of Einstein and continuity equations ~\eqref{equations_motion_xi} is highly non-linear. In particular supersonic deflagration seems to be the mode that mostly suffers from numerical error, as the compression wave can become very narrow and not well resolved (see the central panels of Figure~\ref{fig:mink}). 
Hybrid solutions indeed have an additional source of error that come from the fact that, as explained in Section~\ref{sec:nm_hybrids}, the inward integration from the wall starts at $v_- =1/3$, and a small displacement is needed to leave the fixed point.

\subsection{Energy redistribution}
As the transition proceeds and the bubble expands, part of the energy difference between the symmetric and broken phase is converted into kinetic energy of the fluid around the bubble \cite{Steinhardt:1981ct}. The fluid motion turns on and enhances shear stress perturbations, and becomes a source for gravitational radiation \cite{Witten:1984rs}. The energy budget of the phase transition is crucial to quantify the power spectrum of gravitational waves \cite{Hindmarsh:2013xza, Hindmarsh:2016lnk, Hindmarsh:2019phv}. 

The total amount of energy that was contained in the volume occupied by the bubble before the transition happened is
\begin{equation}
    E_b = \int_0^{R_w}  4\pi e_F R^2 dR = \frac{t}{3}\Omega_F(\xi_w)\xi_w,
\end{equation}
where $e_F$ is the homogeneous and isotropic energy density of the outer FLRW Universe. 

During the transition, the energy density of the fluid as measured by constant-$\xi$ observers $U^\mu$~\eqref{def:U_mu_obs}, that is observers that move with speed $u$ relative to the fluid, is $T_{\mu\nu}U^\mu U^\nu$. We extract the kinetic part by subtracting the energy density in the rest frame of the fluid from the total energy density~\cite{Cai:2018teh, Hindmarsh:2020hop}
\begin{equation}
    e_{K} \equiv T_{\mu\nu}U^\mu U^\nu - T_{\mu\nu}u^\mu u^\nu,
\end{equation}
and we get
\begin{equation}\label{eq:e_K}
    e_K = wu^2\gamma_u^2.
\end{equation}
The kinetic energy $E_K$ of the fluid is then calculated by integrating~\eqref{eq:e_K} over the volume of the bubble
\begin{equation}
    E_K = t\int d\xi \, (1+\omega)\Omega u^2\gamma_u^2,
\end{equation}
and the ratio 
\begin{equation}\label{def:kinetic_fraction}
    \mathcal{K} \equiv \frac{E_K}{E_b} = \frac{3\int d\xi \, (1+\omega)\Omega u^2\gamma_u^2}{\xi_w\Omega_F(\xi_w)}
\end{equation}
measures the amount of energy contained in the bubble that has been converted into kinetic energy during the transition.

\begin{figure}
    \centering
    \includegraphics[width=\textwidth]{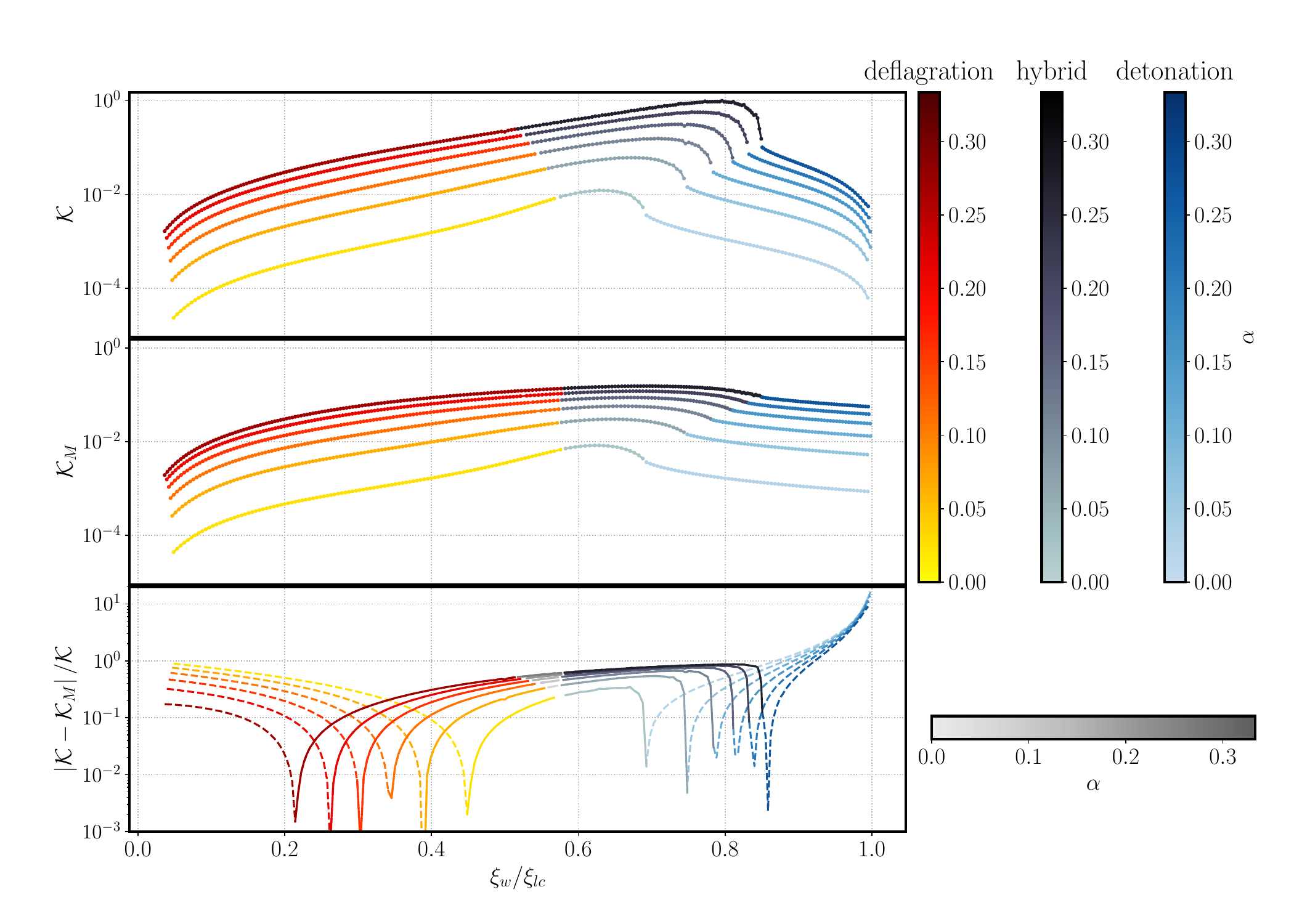}
    \caption{Kinetic energy fraction $\mathcal{K}$ against the fractional distance of the wall from the light cone $\xi_w/\xi_{lc}$ in the gravitating solutions (top panel) and for the case of bubbles expanding on a Minkowski flat background (central panel) with constant sound speed EOS~\eqref{def:eos}~\eqref{def:omega}, $\mathcal{K}_M$. Absolute relative difference between the two cases above in the bottom panel. Dashed lines are used when the relative difference is negative, and solid line when it is positive. Some of the gravitating solutions in the hybrid regime belong to the deflagration regime in the Minkowski case for the same value of $\xi_w/\xi_{lc}$. For these solutions we use a grey colormap in the bottom plot.}
    \label{fig:kinetic_energy}
\end{figure}
We show the kinetic energy fraction obtained from the gravitating solutions in the top panel of Figure~\ref{fig:kinetic_energy}. We compare these results with the kinetic energy fraction that we obtain from a single bubble with the same EOS~\eqref{def:eos}~\eqref{def:omega} expanding on a Minkowski flat background (central panel). For the latter case, we made use of the publicly available code in Ref.~\cite{Giese:2020znk} to get the efficiency factor $\kappa_v = 4 e_K/\Delta\theta$, which is related to the kinetic energy fraction through~\cite{Hindmarsh:2019phv, Hindmarsh:2020hop}
\begin{equation}
    \mathcal{K} = \frac{\kappa \alpha_T}{1+\alpha_T+\delta_n},
\end{equation}
where $\delta_n \equiv 4\theta_b/(3w_s)$ is always zero in our case because $\theta_b = e_b(1-3\omega_-) = 0$, since $\omega_- =1/3$. 

We point out that the kinetic energy fraction in the gravitating solutions exceeds the one in the Minkowski solutions for the same transition strength $\alpha$ in the case of hybrids and fast deflagrations, while the Minkowski solutions are more efficient in the other cases. The relative difference between the two solutions (bottom panel of Figure~\ref{fig:kinetic_energy}) reaches order unity in the hybrid and deflagration regime, while it grows up to almost order $\mathcal{O}(10)$ for the fastest detonations. This might have important consequences in the generation of shear stress in the fluid and, consequently, in the sourcing of gravitational waves. We notice further that the gravitating solutions always have a visible peak of $\mathcal{K}$ in the hybrid regime, while the same peak in the Minkowski solutions becomes more and more broad with increasing $\alpha$. The detonation sector shows a decrease of $\mathcal{K}$ with the wall position $\xi_w/\xi_{lc}$ in both cases, but the curves are much steeper in the gravitating solutions.

\section{Conclusions}\label{sec:conclusions}
We presented new self-similar solutions of the general relativistic hydrodynamic equations that describe the expansion of bubbles of the broken phase into the symmetric phase, in the limit that the ratio of the proper wall width and the bubble size go to zero.

To find these solutions, we assumed a simple equation of state compatible with self-similarity.  
The transition must be strongly supercooled, as the bubbles need a long time $t\gg t_n = t(T_n)$ to reach sizes comparable to the Hubble radius. In the simplified equation of state model, the energy density of the metastable phse~\eqref{def:theta} decreases as $T^{-4}$, which is different
from the standard scenario of supercooled phase transitions~\cite{Lewicki:2022pdb, Ellis:2019oqb, Ellis:2020nnr, Baratella:2018pxi, Prokopec:2018tnq, Konstandin:2011dr, Kierkla:2022odc}, where the energy density of the false vacuum is asymptotically constant in time. Nonetheless, we expect that the solutions given some guidance 
on the gravitational effects to be expected in the standard scenario, where large density perturbations $\delta e/e \sim \mathcal{O}(1)$ are also produced.

Relaxing the self-similar assumption requires solving the equations of motion as a system of partial differential equations, and might be a challenging topic for further studies. The self-simular  solutions will be a useful benchmark for such simulations. 

Our solutions show qualitative and quantitative differences respect to the analogous Minkowski solutions. Constant-$\xi$ observers see a local departure from a FLRW fluid $u$ as a rarefaction wave behind the wall that extends to the origin. In the Minkowski solutions this rarefaction wave extends inward only as far as the sound speed in the detonation and supersonic deflagration case, and it is not present at all in deflagrations. As expected from the Minkowski solutions, the supersonic deflagration is the mode with strongest discontinuities in the fluid speed $v$ and energy $\Omega$ at the wall.

Gravitational effects modify the distribution of the kinetic energy of the fluid around the bubble. We found that the kinetic energy fraction $\mathcal{K}$~\eqref{def:kinetic_fraction} is enhanced by gravitational effects in the regime of supersonic deflagrations and fast subsonics deflagration, while detonations and slow deflagrations on a curved spacetime are less efficient than the same solutions on a Minkowski flat spacetime. This will have direct consequences in the estimation of the power spectrum of gravitational waves \cite{Hindmarsh:2013xza, Hindmarsh:2016lnk, Hindmarsh:2019phv}, and might significantly change its amplitude when the relative difference between the two cases is of order $\mathcal{O}(1)$.

In all cases, the bubble shows a deficit in the energy density inside the wall respect to the outside. This fact implies that the spatial sections inside the wall are negatively curved. We quantified the general relativistic effect on the energy deficit inside the wall by looking at the amount of spatial curvature at the centre of the bubble and comparing with a quasi-Newtonian estimate derived by the solutions in Minkowski spacetime. The amplitude of spatial Ricci curvature has a peak in the hybrid regime and is, but the latter case shows larger negative curvature in the subsonic deflagration cases, while the former has larger negative curvature in the detonation regime.

Finally we mention prospects for gravitational wave observations. A simple analysis of the Weyl tensor components shows that spherical symmetry protects the single bubble configuration from sourcing gravitational waves. Primary gravitational waves can instead arise when the spherical symmetry is broken, i.e. by collision of bubbles and sound waves, whose initial conditions are set by the velocity and energy density profiles around the bubble~\cite{Hindmarsh:2016lnk, Hindmarsh:2019phv}.  Gravitational effects change these profiles, which can be expected to lead to modifications in the gravitational wave power spectrum.

We also expect bubbles to produce metric perturbations which source secondary gravitational waves. The source of primary gravitational wave is approximately $\delta T_{ij} \sim H^2 \left(\delta e/e\right)$, while the source of scalar induced gravitational waves can be estimated as 
\begin{equation}
\partial_i\Phi_B\partial_j\Phi_B \sim R_\star^{-2} (H R_\star)^4 (\delta e/e)^2,
\end{equation}
with $\Phi_B$ the scalar Bardeen potential. The relative contribution of scalar induced gravitational waves
\begin{equation}
    \frac{\partial_i\Phi_B \partial_j\Phi_B}{\delta T_{ij}} \sim (H R_\star)^2\left(\frac{\delta e}{e}\right)
\end{equation}
becomes relevant when $R_\star H \sim \mathcal{O}(1)$. A better estimate of the secondary gravitational waves generated by the expanding bubble in slow FOPTs can be achieved by performing a complete analysis of scalar cosmological perturbations~\cite{Sasaki:1982fi}. 

\section*{Acknowledgements}
We thank Chris Byrnes for discussion and comments, and Mika Mäki for sharing pre-release code for comparison of Minkowski space numerical solutions in the constant sound speed model. LG and MH are supported by Academy of Finland grant number 333609.

\appendix
\section{Junction conditions}\label{appendix_jc}
Consider a timelike hypersurface $\Sigma$ of codimension $1$ embedded in a 4-dimensional spacetime $\mathcal{M}$ with metric $g$. $\Sigma$ divides the spacetime in two regions, $\mathcal{M}^+$ outside the surface, and $\mathcal{M}^-$ inside, respectively with metric $g^+$ and $g^-$ and coordinates $x^+$ and $x^-$. In this work we used the notation of~\cite{Tanahashi:2014sma}
\begin{gather}
    \left\{ A \right\}^\pm \equiv A^+ + A^-,\\
    \left[ A \right]^\pm \equiv A^+ - A^-,\\
    \Bar{A} \equiv \frac{1}{2}\left\{ A \right\}^\pm,
\end{gather}
where the suffix means that a quantity $A$ is evaluated just inside ($-$) or just outside ($+$) the hypersurface $\Sigma$. We wish to find the conditions on the metric and energy momentum tensor that ensure that the union of $g^+$ and $g^-$ is a valid solution to the Einstein equations. The problem of junction conditions was studied originally by Israel~\cite{Israel:1966rt} and has been studied widely afterwards in literature using different formalism~\cite{Maeda:1985ye, Tanahashi:2014sma, Deng:2017uwc, 2013rehy.book.....R}. For our purposes, we find convenient to use a similar approach as Poisson~\cite{Poisson:2009pwt}.

First of all, we need to assume that a continuous coordinate system $x^\mu$ (different from $x_\pm^\mu$, that, in principle, might not match in $\Sigma$) exists such that the matching between the two manifolds $\mathcal{M}^\pm$ happens smoothly (without jumps in the metric) at the interface. The existence of this coordinate system is not ensured in  complete generality, but it is ensured when we have high enough symmetries to reduce the degrees of freedom in the metric. This is indeed the case for a $d=4$ space-time with spherical symmetry~\cite{Berezin:1987bc}. Having established the existence of this coordinate system, the idea is to derive the junction conditions in this particularly easy framework, and then project the conditions into the three-dimensional hypersurface in order to get rid of the choice of the particular coordinate system.

Denoting as $g^\pm_{\mu\nu}$ the metric in $\mathcal{M^\pm}$ in coordinates $x^\mu$, we write the metric $g_{\mu\nu}$ in as a distribution valued tensor
\begin{equation}
    g_{\mu\nu} = \Theta(\ell)g^+_{\mu\nu} + \Theta(-\ell)g^-_{\mu\nu}
\end{equation}
with $\Theta$ the Heaviside distribution and $\ell$ the proper distance from the hypersurface\footnote{$\ell$ is taken to be negative inside $\Sigma$ and positive outside.}. Let now $y^a$ be a coordinate system on $\Sigma$. Given the fact that the coordinate system $x^\mu$ is continuous across the hypersurface, the same system $y^i$ can be installed in the two sides of the interface. Let further $n^\mu$ be the orthogonal vector to $\Sigma$ and $e^\mu_i \equiv \partial x^\mu/\partial y^i$ its tangent vectors. The tangent bundle of $\Sigma$ at any point is spanned by a basis of three vielbeins $e_a, \; a=0,2,3$ such that locally the metric induced on the hypersuface is
\begin{equation}\label{eq:induced}
    h_{ab} = e^\mu_a e^\nu_b g_{\mu\nu}.
\end{equation}
The first junction condition states that the induced metric (first fundamental form) is continuous across the interface:
\begin{equation}\label{eq:continuity_h}
    \left[h_{ab} \right]^\pm = 0.
\end{equation}
This condition is necessary in order for the Christoffel symbols of $g_{\mu\nu}$ to be well defined as distributions. Analogously, we can write the energy momentum tensor as
\begin{equation}
    T_{\mu\nu} = \Theta(\ell)T^+_{\mu\nu} + \Theta(-\ell)T^-_{\mu\nu} + \delta(\ell) S_{\mu\nu}
\end{equation}
where $S_{\mu\nu}$ is the part of the energy-momentum tensor that takes into account for a singular distribution of matter at $\Sigma$. The Einstein equations directly provide the second junction condition
\begin{equation}
    S_{ab} = -\dfrac{1}{8\pi G} ([K_{ab}]^\pm - [K]^\pm g_{ab}),
\end{equation}
where the extrinsic curvature 3-tensor (second fundamental form) is defined by
\begin{equation}
     K_{ab} = e^\mu_{\phantom{\mu}a} e^\nu_{\phantom{\nu}b} n_{\mu;\nu}.
\end{equation}
From Gauss-Codazzi equation contracted twice with the orthogonal vector $n^\mu$
\begin{equation}
    G_{\mu\nu}n^\mu n^\nu = \dfrac{1}{2}\left(K^2 -K_{ab}K^{ab} -R^{(3)}\right),
\end{equation}
we have
\begin{equation}\label{JC1}
    S^{ab}\Bar{K}_{ab} = - \dfrac{1}{16\pi G}\left( [K_{ab}K^{ab}]^\pm - [K^2]^\pm \right) = \left[T_{\mu\nu} n^\mu n^\nu \right]^\pm.
\end{equation}
Projecting the Gauss-Codazzi equation onto the orthogonal vector $n^\mu$ once and on the tangent vector $e^\mu_a$ once 
\begin{equation}
    G_{\mu\nu}e^\mu_{\phantom{\mu}a} n^\nu = K_{a\phantom{b};b}^{{\phantom{a}b}} - K_{,a}
\end{equation}
we derive
\begin{equation}\label{JC2}
   S^b_{a;b} = -\dfrac{1}{8\pi G}\Big(\left[K_a^{\phantom{a}b}\right]^\pm_{;b} - \left[K\right]^\pm_{,a} \Big)= - \left[ T_{\mu\nu}e^\mu_{\phantom{\mu}a} n^\nu \right]^\pm,
\end{equation}
where covariant derivatives are computed with the intrinsic metric~\eqref{eq:induced}.

\subsection{Explicit calculation of extrinsic curvature}\label{sec:extrinsic_curvature}
In order to compute the components of the extrinsic curvature $K_{ab}$, we follow the approach of~\cite{Berezin:1987bc}. We introduce a convenient system of coordinates, calculate the extrinsic curvature components, and then re-express the results in terms of coordinate invariant quantities. We first make the coordinate choice
\begin{equation}
    r = R,
\end{equation}
so that the metric in these coordinates is
\begin{equation}
    ds^2 = -a^2dt^2 +b^2dR^2 + R^2d\Omega^2.
\end{equation}
The interface is defined by a constraint $F(x^\mu) =0$ that we assume of the form
\begin{equation}
    F(x^\mu) = R - \mathcal{S}(t) = 0, 
\end{equation}
with $\mathcal{S}(t)$ an arbitrary function of time only. The orthogonal vector is
\begin{equation}
    n_\mu = N\partial_\mu \mathcal{S} = N\left(-\dot{\mathcal{S}}, 1, 0, 0\right), \qquad N= \left[b^{-2} - a^{-2}\dot{\mathcal{S}}^2 \right]^{-1/2}
\end{equation}
with $N$ a normalization factor that ensures $n^\mu n_\mu = 1$. The tangent vectors to the hypersurface $e^\mu_a$ are obtained from the conditions $e^\mu_a n_\mu =0$ and $g_{\mu\nu}e^\mu_0 e^\nu_0 =-1$. We find 
\begin{eqnarray}
    e^\mu_0 &=& \frac{1}{\sqrt{a^2 - b^2\dot{\mathcal{S}}^2}} \left(1, \dot{\mathcal{S}}, 0, 0\right)\\
    e^\mu_2 &=& \left(0,0,1,0\right)\label{eq:theta}\\
    e^\mu_3 &=& \left(0,0,0,1\right)\label{eq:phi}
\end{eqnarray}
Equations~\eqref{eq:theta} and~\eqref{eq:phi} tell us that the angles we use on adapted coordinates are the same angles defined in the global metric. We denote $y^0 \equiv \tau$, since the condition $g_{\mu\nu}e^\mu_0 e^\nu_0 =-1$ gives to $y^0$ the interpretation of the proper time of an observer comoving with the shell. Therefore, the induced line element is
\begin{equation}
    ds_\Sigma^2 = -d\tau^2 + \rho(\tau)^2d\Omega^2,
\end{equation}
with $\rho(\tau) = R_+ = R_-$ as enforced by the first junction condition~\eqref{eq:continuity_h}. Since the quantities defined on the induced metric do not depend on the coordinate choice of the global spacetime $\mathcal{M}$, we would like to rewrite all the basic quantities in term of $\rho$ and $\tau$. With this goal in mind
\begin{equation}
    \dot{\rho} \equiv \frac{d\rho}{d\tau} = \frac{dR}{dt}\frac{dt}{d\tau} = \frac{\dot{\mathcal{S}}}{\sqrt{a^2 - b^2\dot{\mathcal{S}}^2}},
\end{equation}
and we can substitute $\dot{\mathcal{S}}$ for $\dot{\rho}$ in the above equations:
\begin{eqnarray}
    n_\mu &=& \left(-ab\dot{\rho}, b^2\sqrt{{\dot{\rho}}^2 + b^{-2}}, 0, 0\right)\\
    e^\mu_0 &=&  \left(\frac{b}{a} \sqrt{b^{-2} + {\dot{\rho}}^2} , \dot{\rho}, 0, 0 \right)\\
    e^\mu_2 &=& \left(0,0,1,0\right)\\
    e^\mu_3 &=& \left(0,0,0,1\right)
\end{eqnarray}
Now we are ready to compute the extrinsic curvature:
\begin{equation}
    K_{\theta\theta} = \nabla_\theta n_\theta = -\Gamma^\mu_{22} n_\mu = -\Gamma^1_{22} n_1 = \frac{\rho}{b^2}b^2 \sqrt{b^{-2} + \dot{\rho}^2} = \rho \sqrt{b^{-2} + \dot{\rho}^2}
\end{equation}
The $00$ component is easier to compute following reference~\cite{Maeda:1985ye} as
\begin{eqnarray}
    K_{\tau\tau} &=& - n_\mu \dfrac{De^\mu_\tau}{D\tau} = - n_\mu \left( \dfrac{de^\mu_\tau}{d\tau} + \Gamma^\mu_{\rho\sigma} e_\tau^\rho e_\tau^\sigma\right)= \nonumber\\
    &=& - \dfrac{1}{\sqrt{b^{-2} + \dot{\rho}^2}}\left[ \Ddot{\rho} + \left(\frac{b^\prime}{b} +\frac{a^\prime}{a}\right)\dot{\rho}^2 + \frac{a^\prime}{ab^2}\right] - 2 \dfrac{\dot{b}}{a}\dot{\rho}.
\end{eqnarray}
Now we want to re-write these expressions in terms of invariant quantities only. The first invariant is $\rho(\tau)$ and its derivatives. Then we can construct
\begin{subequations}
    \begin{eqnarray}
        \Delta &=& g^{\mu\nu}R_{,\mu}R_{,\nu} = b^{-2}, \\
        \Delta_1 &=& \nabla_\mu (g^{\mu\nu}R_{,\nu}) = \frac{1}{b^2}\left( -\frac{b^\prime}{b} + \frac{a^\prime}{a} + \frac{2}{R} \right),\\
        \Delta_2 &=& g^{\mu\nu}R_{,\mu} \Delta_{,\nu} = -2\frac{b^\prime}{b^5},\\
        \Delta_3 &=& g^{\mu\nu}\Delta_{,\mu}\Delta_{,\nu}  = 4\left[ \left(\frac{b^\prime}{b^4}\right)^2 - \left(\frac{\dot{b}}{ab^3}\right)^2\right],
    \end{eqnarray}
\end{subequations}
where the prime denotes derivatives with respect to $r$, $'= \partial/\partial_r$. 
Hence
\begin{subequations}
    \begin{eqnarray}
    \frac{b^\prime}{b} &=& -\frac{1}{2}\frac{\Delta_2}{\Delta^2}\\
    \frac{a^\prime}{a} &=& \frac{\Delta_1}{\Delta} -\frac{1}{2}\frac{\Delta_2}{\Delta^2} -\frac{2}{\rho}\\
    \frac{\dot{b}}{a} &=& \frac{1}{2\Delta^2}\sqrt{\Delta_2^2 - \Delta\Delta_3}
\end{eqnarray}
\end{subequations}
and the extrinsic curvature components become
\begin{eqnarray}
    K^\theta_{\phantom{\theta}\theta} &=& \frac{1}{\rho} \sqrt{\dot{\rho}^2 + \Delta} \\
    K^\tau_{\phantom{\tau}\tau} &=& - \left[ \dfrac{\Ddot{\rho} +\Delta_2/2\Delta}{\sqrt{\dot{\rho}^2 + \Delta}} + \left(\frac{\Delta_1}{\Delta} -\frac{\Delta_2}{\Delta^2} -\frac{2}{\rho} \right) \sqrt{\dot{\rho}^2 + \Delta} + \frac{\dot{\rho}}{\Delta^2} \sqrt{\Delta^2_2 - \Delta\Delta_3} \right].
\end{eqnarray}
Using the Einstein equations one can simplify the time component (see reference~\cite{Berezin:1987bc} for a derivation) to
\begin{eqnarray}
    K^\theta_{\phantom{\theta}\theta} &=& \frac{1}{\rho} \sqrt{\dot{\rho}^2 + \Delta} \\
    K^\tau_{\phantom{\tau}\tau} &=& - \frac{1}{\sqrt{\dot{\rho}^2 +\Delta}}\left[ \Ddot{\rho} + \dfrac{1-\Delta}{2\rho} + 4\pi G \rho T^n_n \right].
\end{eqnarray}
Therefore, given that $\rho(\tau)$ and its derivatives are continuous, the angular component of the junction condition~\eqref{junc_K} demands
\begin{equation}\label{eq:delta_pm}
    \left[\Delta\right]^\pm =0
\end{equation}
The same condition~\eqref{junc_K} for the time component for $\left[K_{\tau\tau}\right]^\pm = 0$ is already encoded in the conservation of the projected energy momentum tensor~\eqref{eq:T_nn} $\left[T^n_n\right]^\pm = 0$; thus it does not bring any additional information. Now we notice that equation~\eqref{eq:delta_pm} is a coordinate invariant statement, that is valid in any coordinate system. We can then specify our system~\eqref{eq:metric} and find
\begin{equation}
    \Delta = -U^2 + \Gamma^2.
\end{equation}
Using the definitions~\eqref{def:delta} and~\eqref{def:omega-phi} of the dimensionless gravitational potential $\Phi$, we finally recover the result of equation~\eqref{eq:junction_phi}:
\begin{equation}
    \left[\Phi\right]^\pm =0.
\end{equation}

\bibliographystyle{JHEP}
\bibliography{references}
\end{document}